\documentclass[twocolumn]{aastex631}

\usepackage{graphicx}
\let\tablenum\relax
\usepackage{pstricks-add}
\usepackage{epsf}
\usepackage{lipsum}% http://ctan.org/pkg/lipsum
\usepackage{hyperref}

\providecommand{\bjdtdb}{\ensuremath{\rm {BJD_{TDB}}}}
\providecommand{\feh}{\ensuremath{\left[{\rm Fe}/{\rm H}\right]}}
\providecommand{\teff}{\ensuremath{T_{\rm eff}}}

\providecommand{\msun}{\ensuremath{\,M_\Sun}}
\providecommand{\rsun}{\ensuremath{\,R_\Sun}}
\providecommand{\lsun}{\ensuremath{\,L_\Sun}}
\providecommand{\mj}{\ensuremath{\,M_{\rm J}}}
\providecommand{\rj}{\ensuremath{\,R_{\rm J}}}

\providecommand{\fave}{\langle F \rangle}
\providecommand{\fluxcgs}{10$^9$ erg s$^{-1}$ cm$^{-2}$}
\providecommand{\kms}{\ensuremath{\,\rm{\,km\,s}^{-1}}}

\providecommand{\vsini}{\ensuremath{v\sin i}}

\shorttitle{HIP 61637 B}
\shortauthors{Ephremidze et al.}

\graphicspath{{./}{figures/}}

\begin{document}

\title{HIP 61637 b: a TESS Brown Dwarf in a Near-circular Orbit around a Massive A-type Star}
%\footnote{Released on March, 1st, 2021}}

\author[0000-0002-9578-3129]{Nino Ephremidze}
\affiliation{Harvard University, Cambridge, MA 02138, USA}

\author[0000-0001-9911-7388]{David W. Latham}
\affiliation{Center for Astrophysics $|$ Harvard $\&$ Smithsonian, 60 Garden Street, Cambridge, MA 02138, USA}

\author[0009-0005-7108-9502]{Perry Berlind}
\affiliation{Center for Astrophysics $|$ Harvard $\&$ Smithsonian, 60 Garden Street, Cambridge, MA 02138, USA}

\author[0000-0001-6637-5401]{Allyson Bieryla}
\affiliation{Center for Astrophysics $|$ Harvard $\&$ Smithsonian, 60 Garden Street, Cambridge, MA 02138, USA}

\author[0000-0002-2830-5661]{Michael L. Calkins}
\affiliation{Center for Astrophysics $|$ Harvard $\&$ Smithsonian, 60 Garden Street, Cambridge, MA 02138, USA}

\author[0000-0002-9789-5474]{Gilbert A. Esquerdo}
\affiliation{Center for Astrophysics $|$ Harvard $\&$ Smithsonian, 60 Garden Street, Cambridge, MA 02138, USA}

\author[0000-0002-8964-8377]{Samuel N. Quinn}
\affiliation{Center for Astrophysics $|$ Harvard $\&$ Smithsonian, 60 Garden Street, Cambridge, MA 02138, USA}

\author[0000-0001-8812-0565]{Joseph E. Rodriguez}
\affiliation{Center for Data Intensive and Time Domain Astronomy, Department of Physics and Astronomy, Michigan State University, East Lansing, MI 48824, USA}

\author[0000-0002-0701-4005]{Noah Vowell}
\affiliation{Center for Data Intensive and Time Domain Astronomy, Department of Physics and Astronomy, Michigan State University, East Lansing, MI 48824, USA}

\author[0000-0002-7382-0160]{Jack Schulte}
\affiliation{Center for Data Intensive and Time Domain Astronomy, Department of Physics and Astronomy, Michigan State University, East Lansing, MI 48824, USA}

\author[0000-0002-4047-4724]{Hugh P. Osborn}
\affiliation{Center for Space and Habitability, University of Bern, Gesellschaftsstrasse 6, 3012 Bern, Switzerland}
\affiliation{ETH Zurich, Department of Physics, Wolfgang-Pauli-Strasse 2, 8093 Zurich, Switzerland}
%\collaboration{6}{(AAS Journals Data Editors)}

%\author{Butler Burton}
%\affiliation{Leiden University}
%\affiliation{AAS Journals Associate Editor-in-Chief}

\begin{abstract}

We present the characterization of HIP 61637 b (TOI-5401 b), a brown dwarf discovered by TESS to transit an A-type star. HIP 61637 is the most massive and the brightest star known to host a transiting brown dwarf to date.  The companion lies in the middle of the ``brown dwarf desert". We perform a joint analysis of light curves from NASA’s TESS mission and our high-resolution spectroscopy from the Tillinghast Reflector Echelle Spectrograph. We determine that HIP 61637 b has a radius of $R_{BD} = 1.149^{+0.049}_{-0.038}$ $R_J$, a mass of $M_{BD} = 47.8^{+1.5}_{-1.4}$ $M_J$, and transits its host star every $6.829104 \pm 0.000011$ days in a near-circular orbit ($e = 0.054 \pm 0.013$). The host star has a mass of $2.86\pm 0.12\,M_\odot$, a radius of $4.33 \pm 0.17\, R_\odot$, and an effective temperature of $T_{\text{eff}} = 9180^{+240}_{-230}$ K. We find that the host is nearing the end of its time on the main sequence and has begun to evolve, allowing for a precise age estimation of $396 \pm 46$ Myr for the system using stellar evolution models. This adds an important data point to the handful of well-characterized transiting brown dwarfs with reliable age estimates, allowing us to test the latest substellar evolution models. Theory of tidal evolution predicts that tidal dissipation mechanisms have circularized the orbit, consistent with the observed near-zero eccentricity.
\end{abstract}

\keywords{Brown dwarfs (185); Spectroscopy (1558); Radial Velocity (1332); Transit Photometry (1709); Stellar Ages (1581); }

\section{\textbf{Introduction}} \label{sec:intro}
Substellar objects that form like stars by the gravitational collapse of interstellar clouds of gas and dust but have masses and temperatures that are insufficient to ignite hydrogen fusion in their cores (75-80 $M_J$, \citealt{Bar02}) are doomed to cool down for the rest of their lifetimes. If they are born massive enough to fuse deuterium (11-16 $M_J$, \citealt{Spi11}), they have historically been called brown dwarfs (BDs), but deuterium is too rare to last very long for its fusion (\citealt{Bur01}). Objects that form by accretion in circumstellar disks of gas and dust have traditionally been called planets, unless they accrete enough mass to fuse deuterium for a portion of their lifetimes, after which they will begin to cool, aside from heating effects due to radioactive decay, stellar irradiation, or tidal interactions with a host star in a tight orbit.  

The traditional definition of BDs based on the mass limits between deuterium and hydrogen fusion therefore encompasses objects formed by at least two very different processes. This was a reasonable choice in the days when detailed characterization of these objects was not possible. This is now changing, largely due to NASA's Transiting Exoplanet Survey Satellite (TESS) mission (\citealt{Rick15}), which is identifying dozens of transiting BDs in relatively short-period orbits that are well suited for radius and mass determinations with transit photometry and radial velocity (RV) observations. There is now some hope that we can distinguish whether a particular transiting BD formed like a planet or like a star: perhaps from demographic studies of the orbits and other characteristics of large samples of BDs, or beyond that, perhaps from the studies of their atmospheres with missions such as JWST, Ariel Mission, and the next generation of giant telescopes, which may be able to reveal whether the BD formed from material in a circumstellar disk based on its atmospheric composition compared to that of the host star.

A key to such studies may be to cover a wide range of host star parameters for transiting BDs, finding young systems with well-determined ages, and uncovering a sizable population in the sparsely populated intermediate-mass region of 40-50 $M_J$ (the so-called ``brown dwarf desert", \citealt{Mar00, Gre06}) where the formation mechanisms are the most ambiguous between star- and planet-like scenarios. This paper presents the confirmation and characterization of such a BD transiting a hot subgiant star. The system is interesting in several ways: HIP 61637 is the most massive and the brightest host star to date with a transiting BD companion, the companion falls in the middle of the BD desert, and the host star is nearing the end of its main-sequence lifetime and beginning to evolve to the subgiant branch, allowing for a relatively well-constrained age determination from stellar evolution tracks.

\begin{deluxetable}{llcc}
\tablecaption{Literature and Measured Properties for HIP 61637 (= TOI-5401, TIC 397058826, HD 109860)}
\tablewidth{0pt}
\tabletypesize{\scriptsize}
\setlength{\tabcolsep}{2pt} % Reduce column separation
\tablehead{
\colhead{\textbf{Parameter}} &
\colhead{\textbf{Description}} &
\colhead{\textbf{Value}} &
\colhead{\textbf{Reference}}
}
\startdata
$\alpha_{J2000}$\dotfill & Right Ascension\dotfill & 12:38:04.38 & a\\
$\delta_{J2000}$\dotfill & Declination\dotfill & +03:16:56.62 & a\\
G\dotfill & Gaia G mag\dotfill & $6.32 \pm 0.02$ & a\\
$B_{\text{P}}-R_{\text{P}}$\dotfill & Gaia $B_{\text{P}}-R_{\text{P}}$ mean mag\dotfill & $0.03\pm 0.02$ & a\\
T\dotfill & TESS mag\dotfill & $6.330 \pm 0.006$ & b\\
 & & & \\
$B_{\text{T}}$\dotfill & Tycho $B_{\text{T}}$ mag\dotfill & $6.336 \pm 0.024$ & b\\
$V_{\text{T}}$\dotfill & Tycho $V_{\text{T}}$ mag\dotfill & $6.329 \pm 0.023$ & b\\
 & & & \\
J\dotfill & 2MASS J mag\dotfill & $6.257 \pm 0.02$ & b\\
H\dotfill & 2MASS H mag\dotfill & $6.301 \pm 0.031$ & b\\
$K_{\text{S}}$\dotfill & 2MASS $K_{\text{S}}$ mag\dotfill & $6.231 \pm 0.021$ & b\\
 & & & \\
WISE1\dotfill & WISE1 mag\dotfill & $6.234 \pm 0.096$ & b\\
WISE2\dotfill & WISE2 mag\dotfill & $6.121 \pm 0.029$ & b\\
WISE3\dotfill & WISE3 mag\dotfill & $6.296 \pm 0.016$ & b\\
WISE4\dotfill & WISE4 mag\dotfill & $6.251 \pm 0.061$ & b\\
 & & & \\
$\mu_{\alpha}$\dotfill & Gaia DR3 proper motion\dotfill & $-30.58 \pm 0.15$ & a\\
 & in RA (mas yr$^{-1}$) & & \\
$\mu_{\delta}$\dotfill & Gaia DR3 proper motion\dotfill & $-11.55 \pm 0.13$ & a\\
 & in Dec (mas yr$^{-1}$) & & \\
 & & & \\
$\pi^{\dag}$\dotfill & Gaia DR3 Parallax (mas)\dotfill & $4.62 \pm 0.05$ & a\\
\enddata
\tablecomments{The uncertainties of the photometry have a systematic error floor applied.\\
RA and Dec are in epoch J2000. \\ 
References: ${}^{\text{a}}$ \citealt{Gai23}, ${}^{\text{b}}$ \citealt{Sta18}.}
\end{deluxetable}

\section{\textbf{Observations}}\label{sec:style}

We obtained photometric and spectroscopic observations of HIP 61637, allowing us to combine transit and RV techniques to characterize the system. We relied on NASA's TESS (\citealt{Rick15}) for transit photometry, which observed 7 full transits of the system spread among 3 different Sectors. The large radius, and consequently the long transit duration (9.8 hours), of HIP 61637 makes it nearly impossible to document full transit events from most ground-based facilities, so we did not undertake any follow-up photometric observations. We obtained RV measurements using the Tillinghast Reflector Echelle Spectrograph (TRES) on the 1.5-m telescope at the Fred Lawrence Whipple Observatory on Mount Hopkins, Arizona.

\begin{figure}
\gridline{\fig{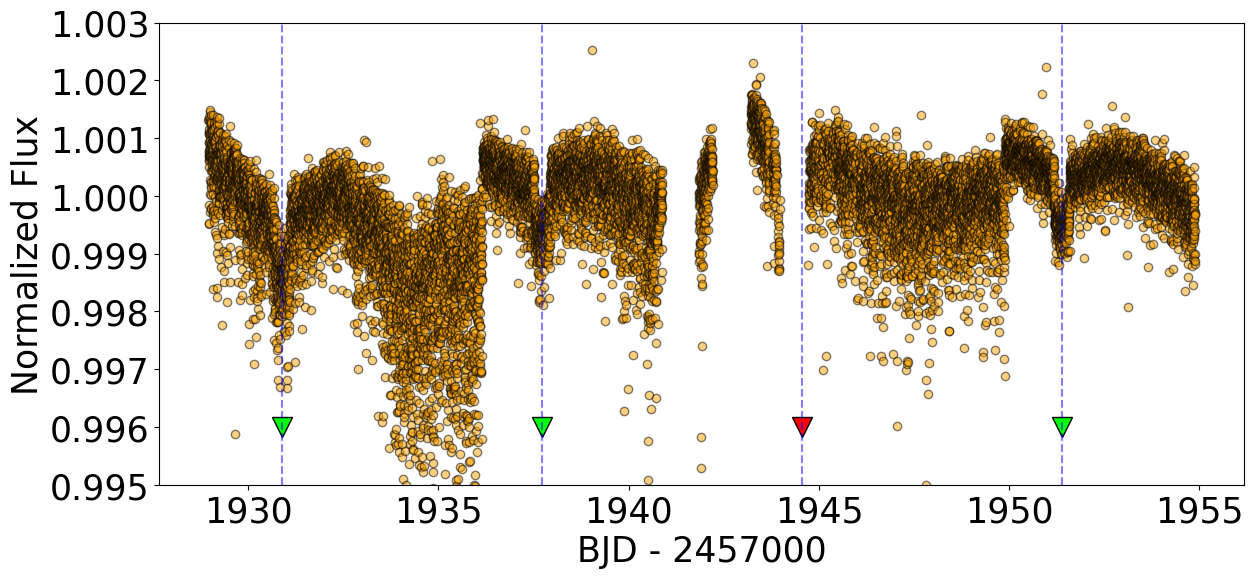}{0.47\textwidth}{Sector 23}}
\gridline{\fig{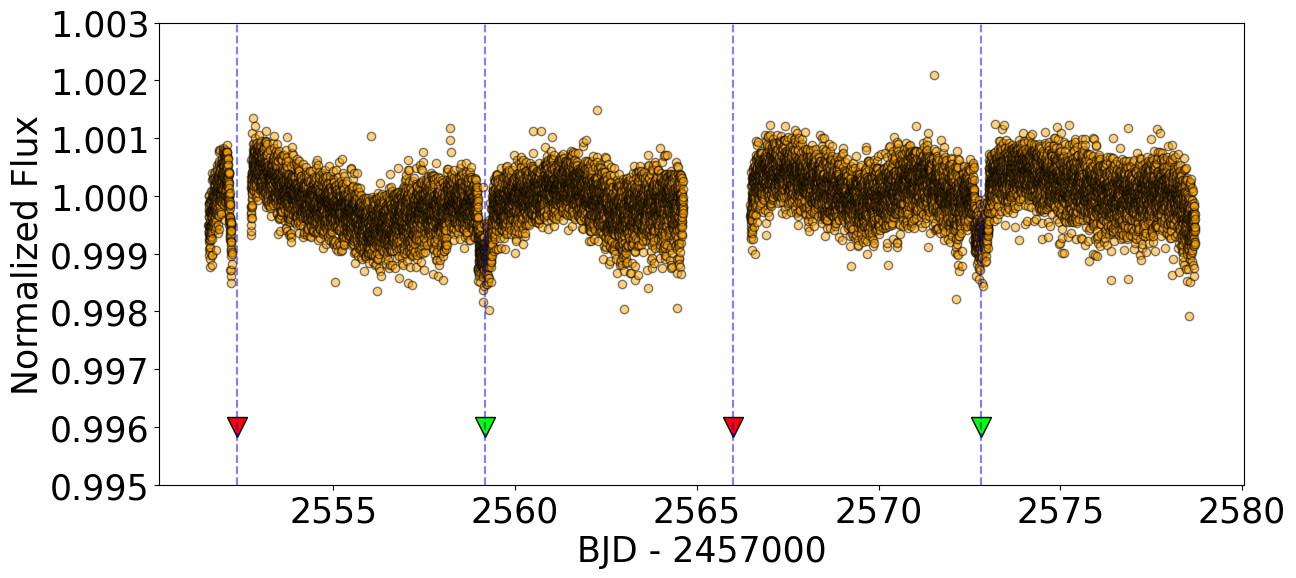}{0.47\textwidth}{Sector 46}}
\gridline{\fig{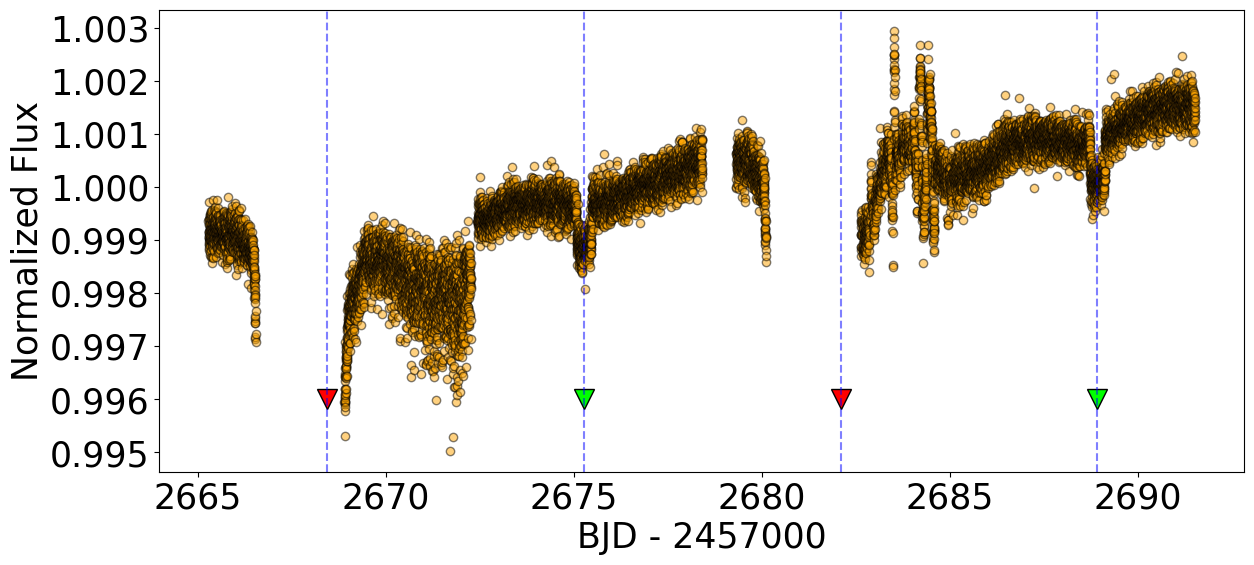}{0.47\textwidth}{Sector 50}}
\caption{Raw light curves of HIP 61637 from TESS, observed in sectors 23, 46, and 50. The flux has been normalized by median deviation. Transit mid-times are marked with dashed vertical lines and triangles. Green triangles indicate complete transit events that were used in our analysis.}
\label{fig:figure1}
\end{figure}

\begin{figure*}
\label{detrending}
\gridline{\fig{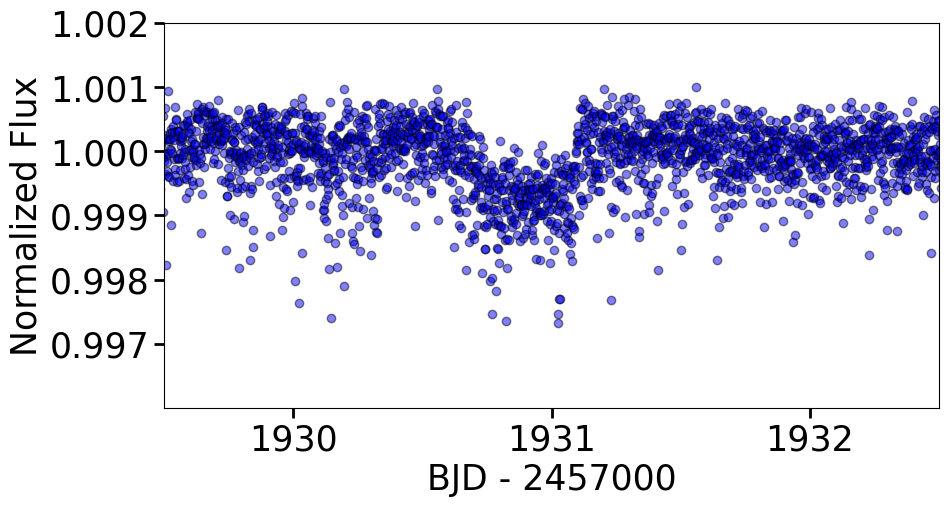}{0.33\textwidth}{(a)}
          \fig{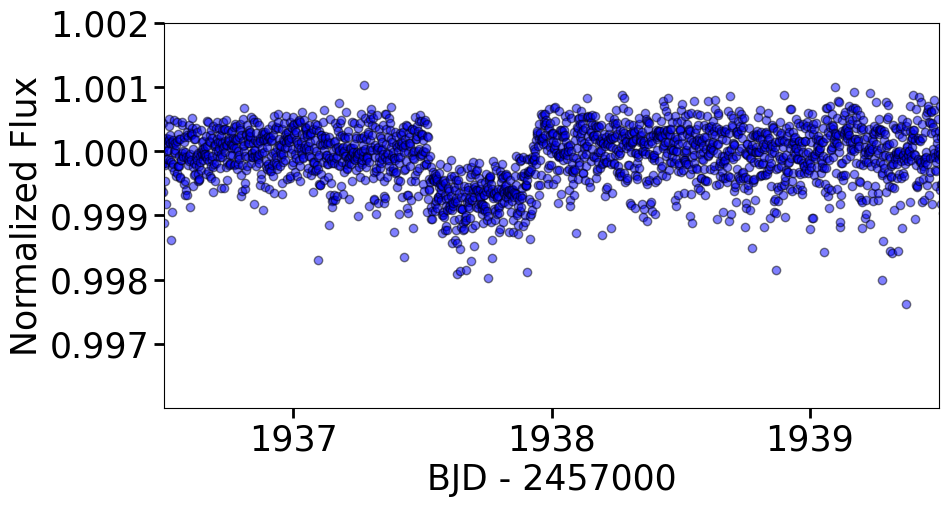}{0.33\textwidth}{(b)}
          \fig{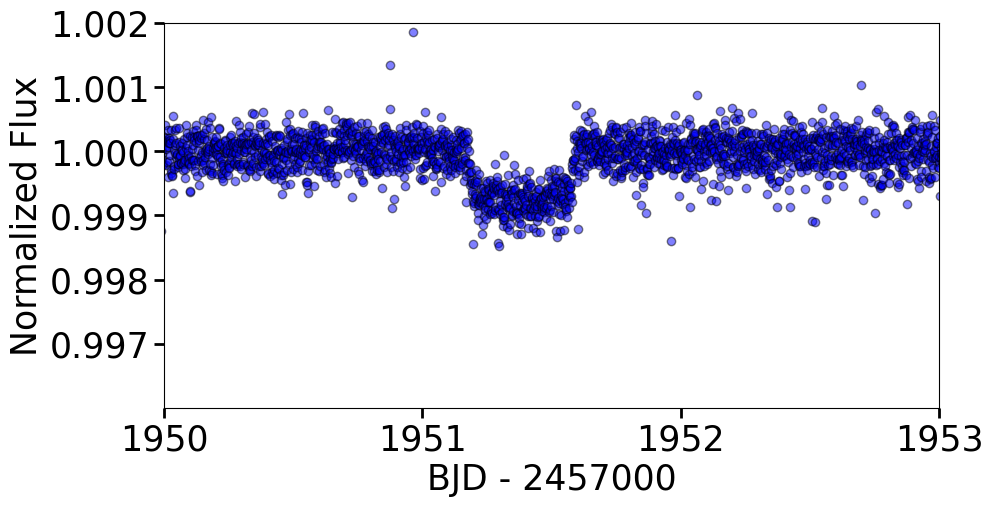}{0.33\textwidth}{(c)}}
\gridline{\fig{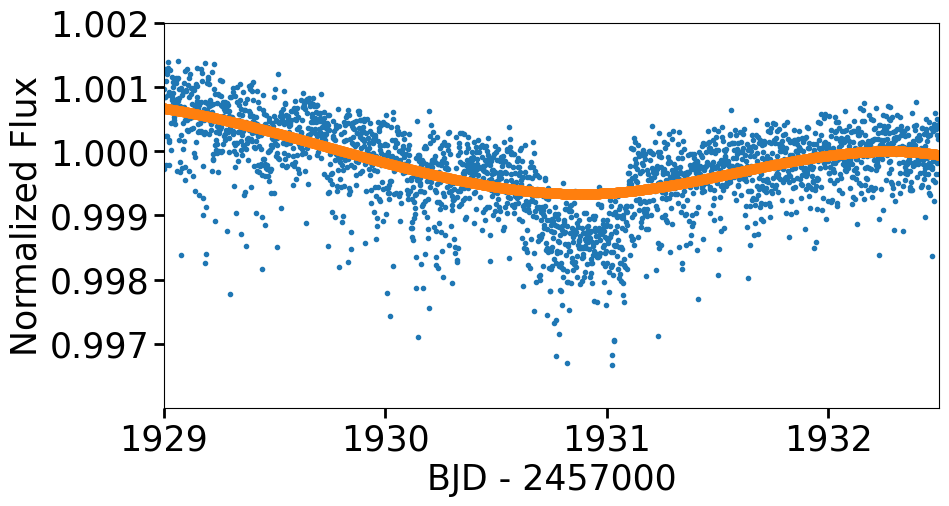}{0.33\textwidth}{(a)}
          \fig{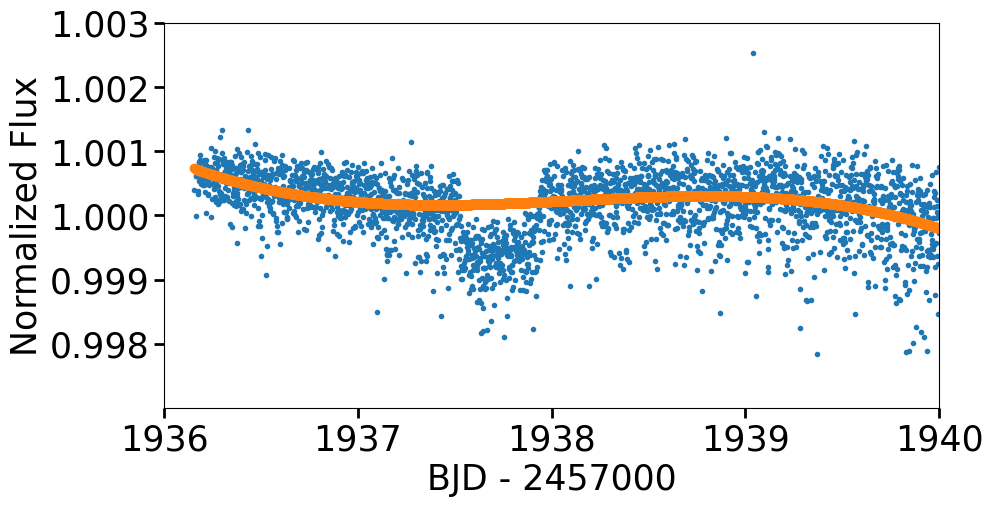}{0.33\textwidth}{(b)}
          \fig{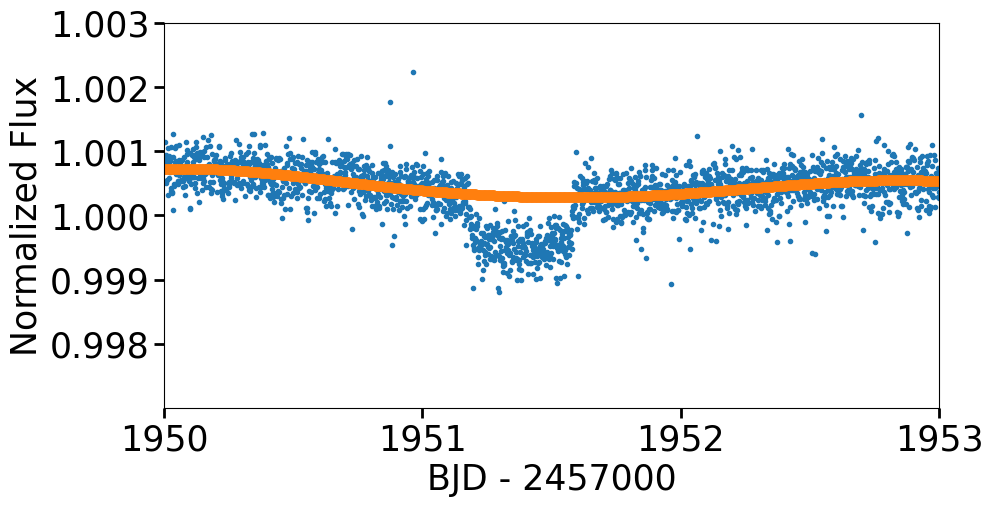}{0.33\textwidth}{(c)}}
\gridline{\fig{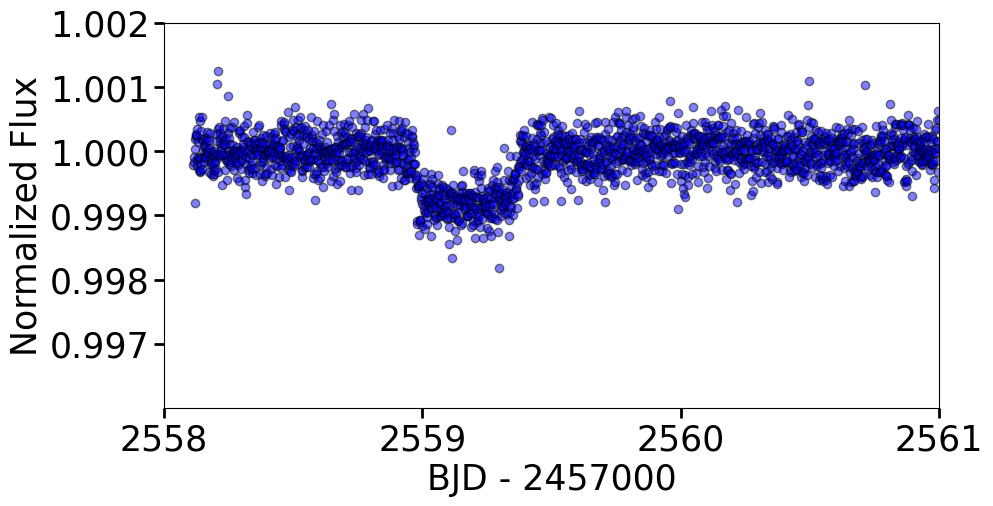}{0.33\textwidth}{(d)}
          \fig{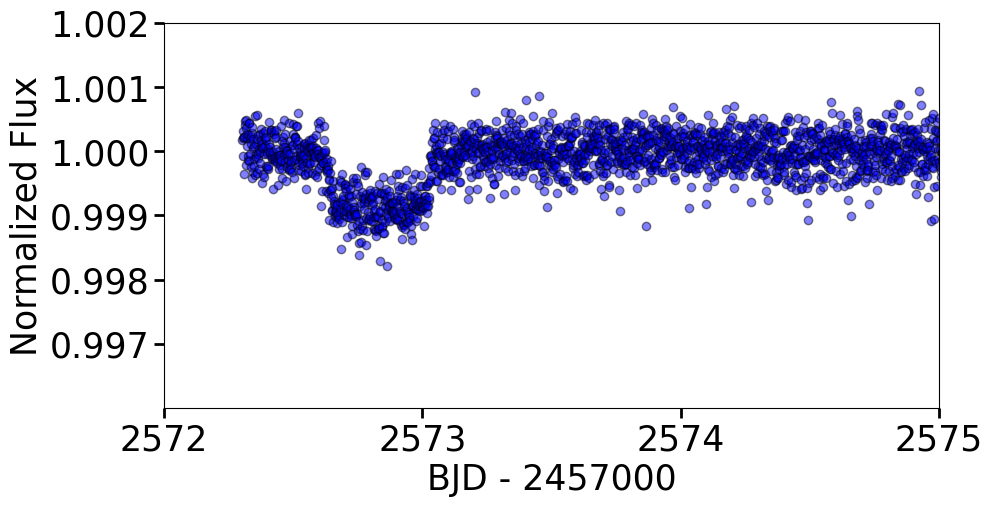}{0.33\textwidth}{(e)}
          \fig{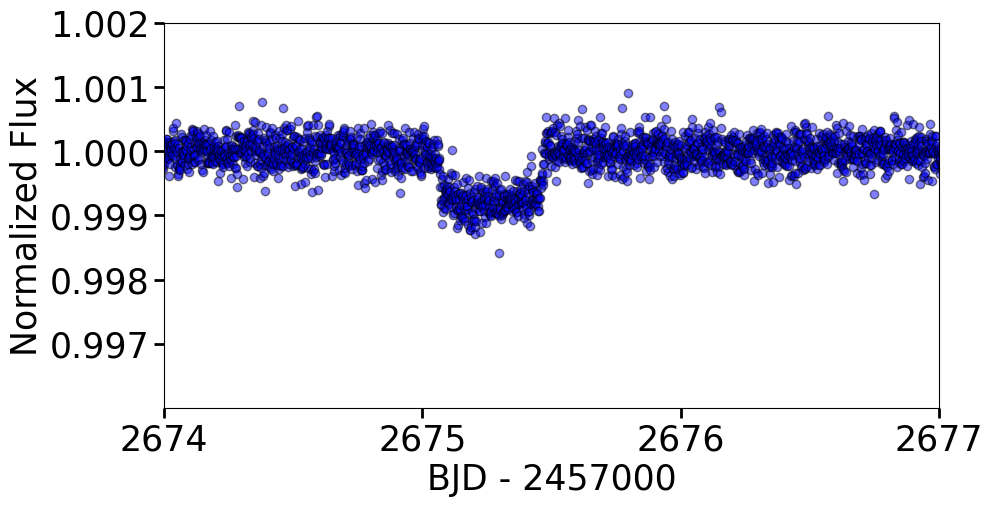}{0.33\textwidth}{(f)}}
\gridline{\fig{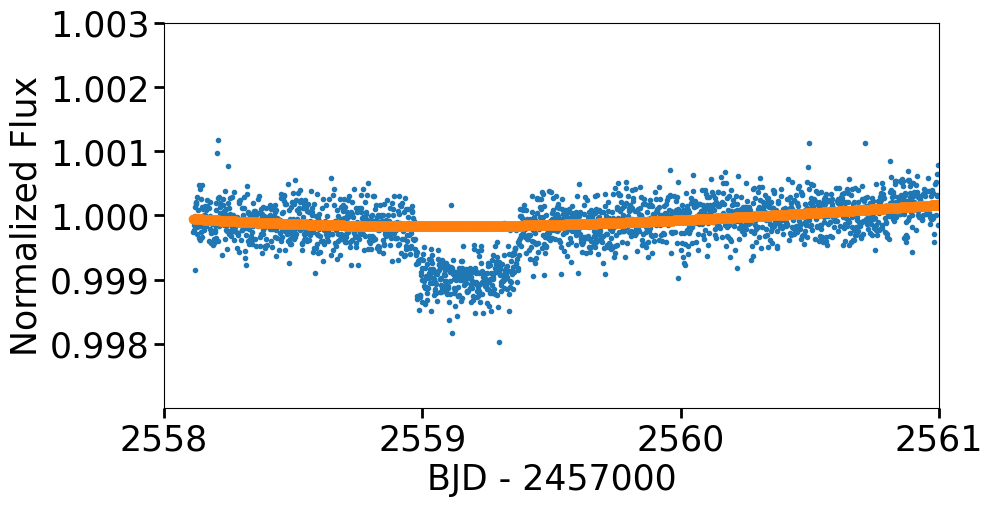}{0.33\textwidth}{(d)}
          \fig{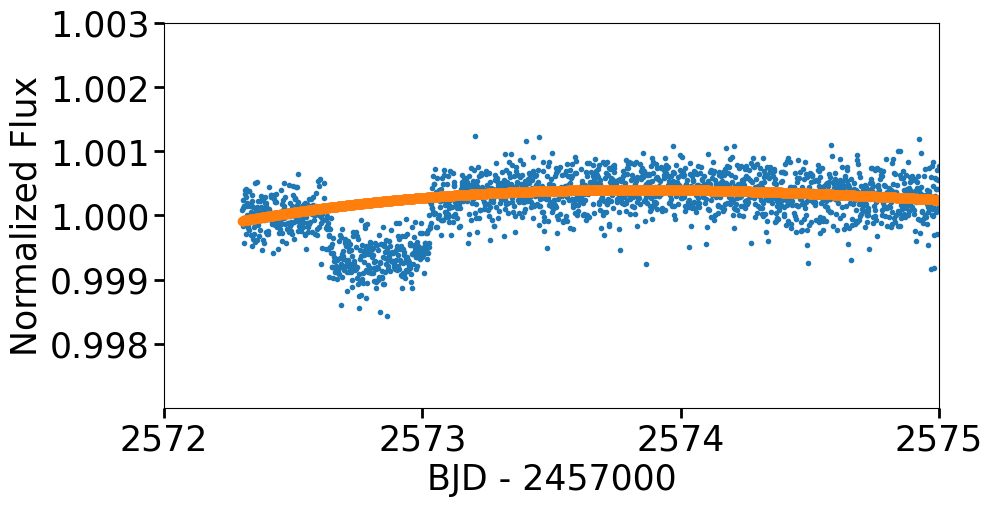}{0.33\textwidth}{(e)}
          \fig{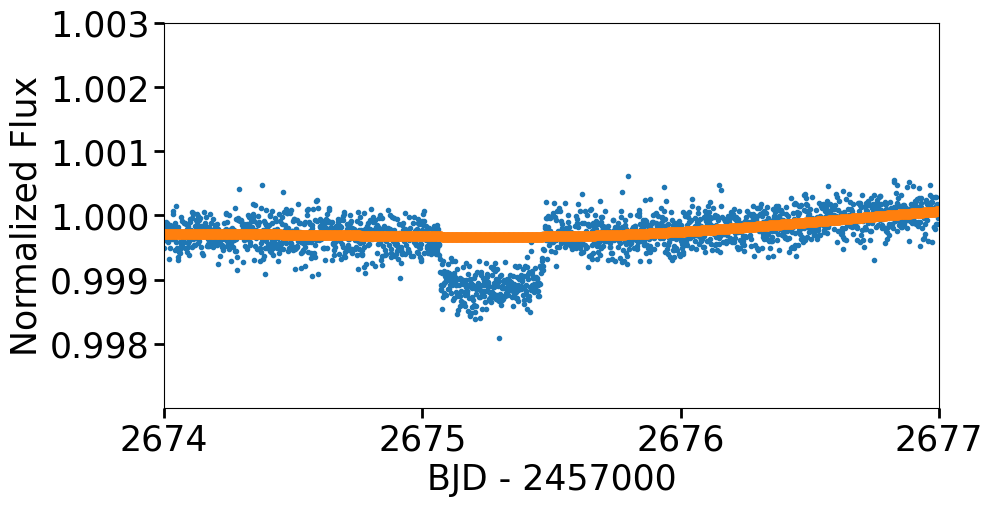}{0.33\textwidth}{(f)}}
\gridline{\fig{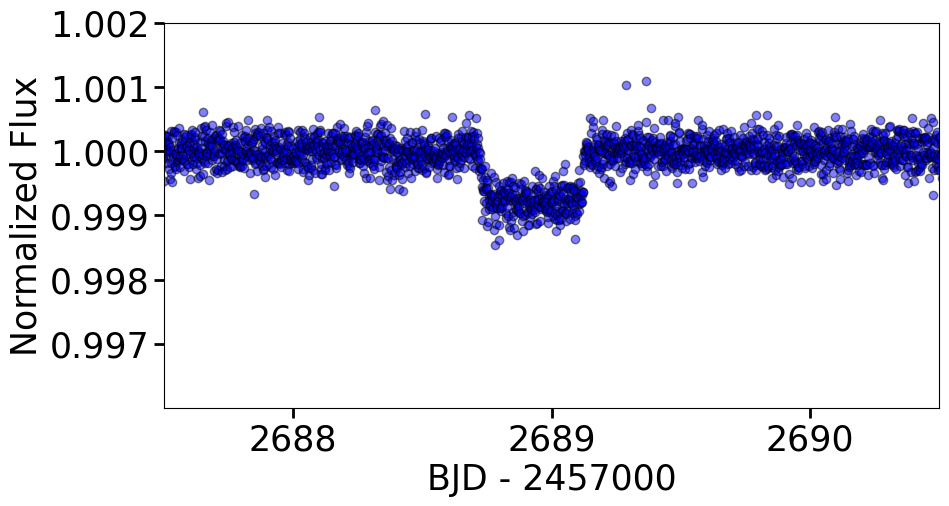}{0.33\textwidth}{(g)}
          \fig{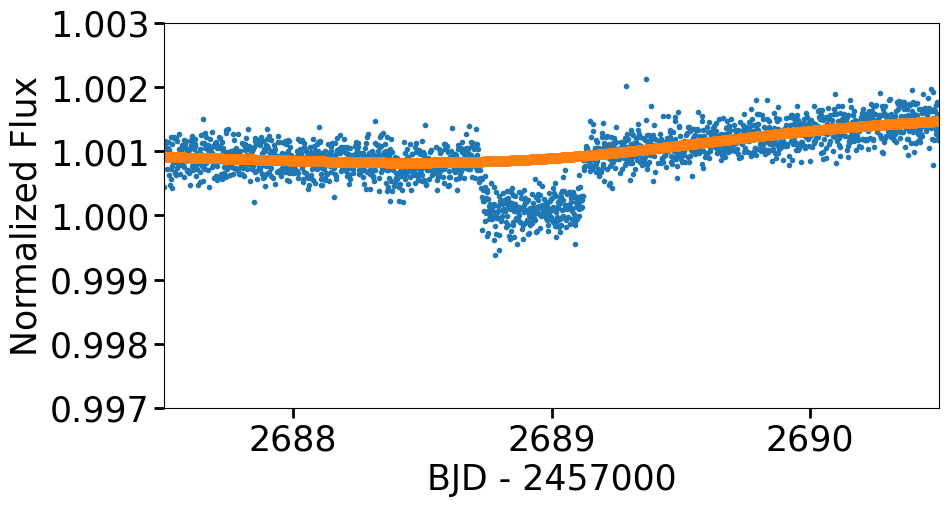}{0.33\textwidth}{(g)}
          }

%\begin{figure*}
%\gridline{\fig{figures/23_1_pro.png}%{0.33\textwidth}{}
%          \fig{figures/23_2_pro.png}{0.33\textwidth}{}
%          \fig{figures/23_3_pro.png}{0.33\textwidth}{}
%          }
%\gridline{\fig{figures/transit_1.png}{0.33\textwidth}{(a)}
%          \fig{figures/transit_2.png}{0.33\textwidth}{(b)}
%          \fig{figures/transit_3.png}{0.33\textwidth}{(c)}
%          }
          
%\gridline{\fig{figures/46_2_pro.png}{0.33\textwidth}{}
%          \fig{figures/46_3_pro.png}{0.33\textwidth}{}
%          \fig{figures/50_1_pro.png}{0.33\textwidth}{}
%          }
          
%\gridline{\fig{figures/transit_4.png}{0.33\textwidth}{(d)}
%          \fig{figures/transit_5.png}{0.33\textwidth}{(e)}
%          \fig{figures/50_1_result.png}{0.33\textwidth}{(f)}
%          }
%\gridline{\fig{figures/50_2_pro.png}{0.33\textwidth}{(g)}
%          \fig{figures/50_2_result.png}{0.33\textwidth}{(g)}}
\caption{
All 7 full transits (labeled a-g), extracted from the 3 sectors of TESS observations. Turquoise plots show the raw light curve data, with the best-fit Keplerspline fits shown with the orange lines. Dark blue plots show the detrended light curves.}
\end{figure*}

\subsection{TESS Photometry}
NASA's TESS (\citealt{Rick15}) is an all-sky survey mission that monitors the brightness of millions of stars with a primary goal of searching for transiting exoplanets. With a total field of view of $24^{\circ} \times 96^{\circ}$, TESS observes the sky one segment at a time and has an observation period of 27.4 days per each of the 26 segments needed to cover almost the entire sky. This makes it more sensitive to systems with orbital periods of 13 or fewer days, so that at least two transits can be captured within its observing period of a particular segment. TESS observes in a broad optical band (600-1000 nm) and achieves a typical precision of better than 50 ppm for bright stars. 
%In its primary mission, which lasted two years, TESS imaged about $75\%$ of the sky, with Full Frame images binned at 30-minute cadence, and a pre-selected 200,000 stars monitored at two-minute cadence. In its extended mission, TESS switched to a 10-minute cadence for Full Frame Images and introduced a 20-second cadence mode for specific stars. Since its launch in 2018, TESS has been responsible for the discovery of more than 230 confirmed extrasolar planets, 10 BDs, and 5,600 TESS Objects of Interest (TOIs).%

TESS first observed HIP 61637 from March 19 to April 13, 2020 in Sector 23 during its primary mission, with a 30-minute cadence. The initial detections of the transits were made by the MIT Quick Look Pipeline (\citealt{QLP1}, \citealt{QLP2}, \citealt{QLP3}, \citealt{QLP4}). TESS observed HIP 61637 again in Sector 46, beginning December 3, 2021, and in Sector 50, beginning March 26, 2022, at a 2-minute cadence. 

We obtained the TESS light curves from the Presearch Data Conditioning Simple Aperture Photometry (PDCSAP; \citealt{Stu12, Stu14, Smi12}) flux produced by the TESS Science Processing Operations Center at NASA Ames, available on the Mikulski Archive for Space Telescopes (MAST)\footnote{\url{https://mast.stsci.edu/portal/Mashup/Clients/Mast/Portal.html}}. We normalized the transit photometry data from the three TESS sectors to produce the raw light curves shown in Figure \ref{fig:figure1}. To remove variations in brightness caused by stellar and/or instrumental variability from the light curves, we applied B-spline detrending using \texttt{Keplersplinev2} software (\citealt{Van2024}). We independently processed sections of the light curve data between momentum dumps of each TESS sector to prevent fitting variability caused by the instrument's operation. Figure \ref{detrending} shows the best-fit splines overplotted on raw TESS data and the detrended light curves for each transit. All seven of the full TESS transits were used in our analysis.

\subsection{TRES Spectroscopy}

\begin{figure}[t!]
    \centering
    \includegraphics[width=\columnwidth]{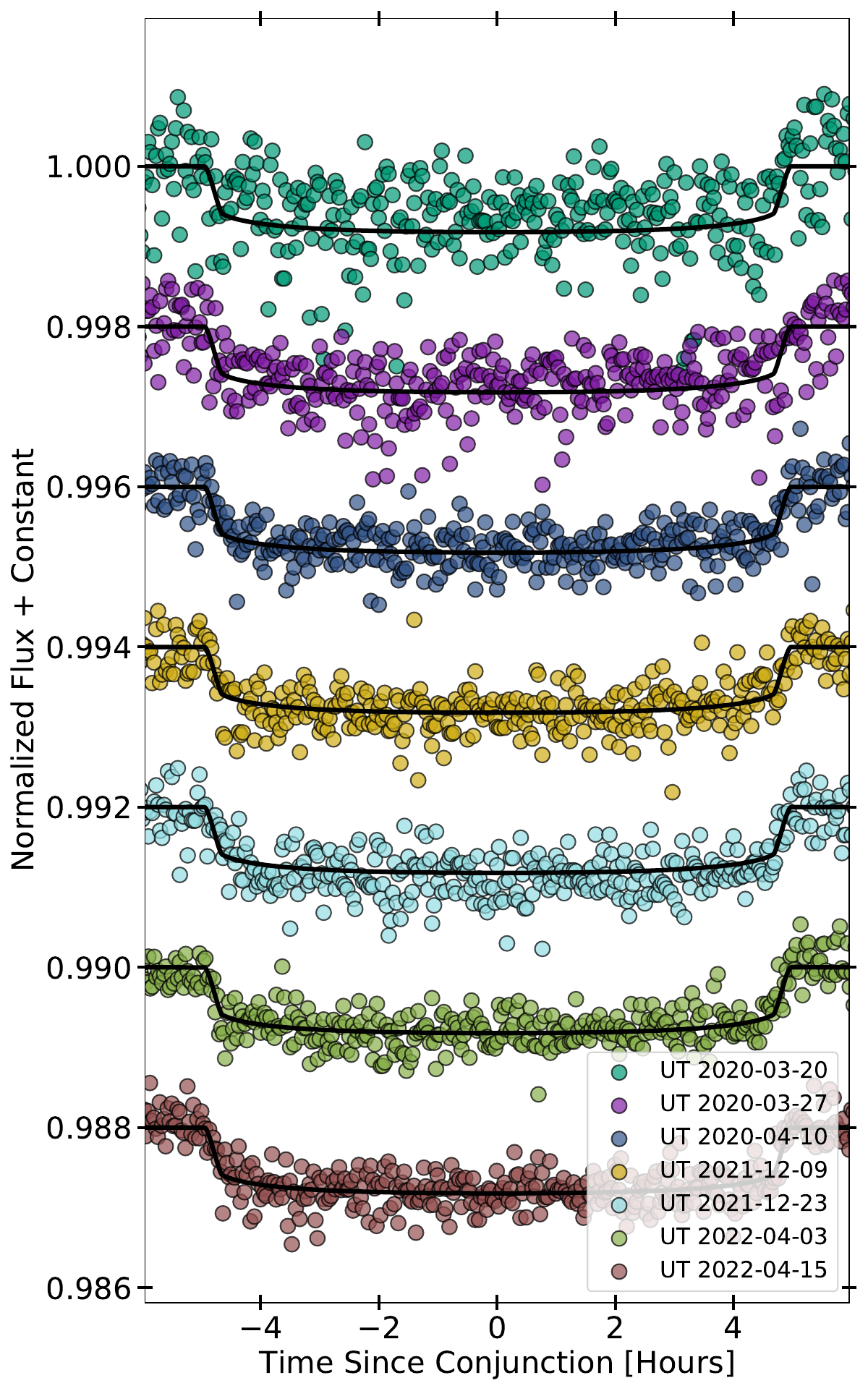}
    \caption{
    The best-fit transit models of TESS lightcurve data from analysis with \texttt{EXOFASTv2} software.
    \label{fig:transits}}
\end{figure}

\begin{figure}[t!]
    \centering
    \includegraphics[width=\columnwidth]{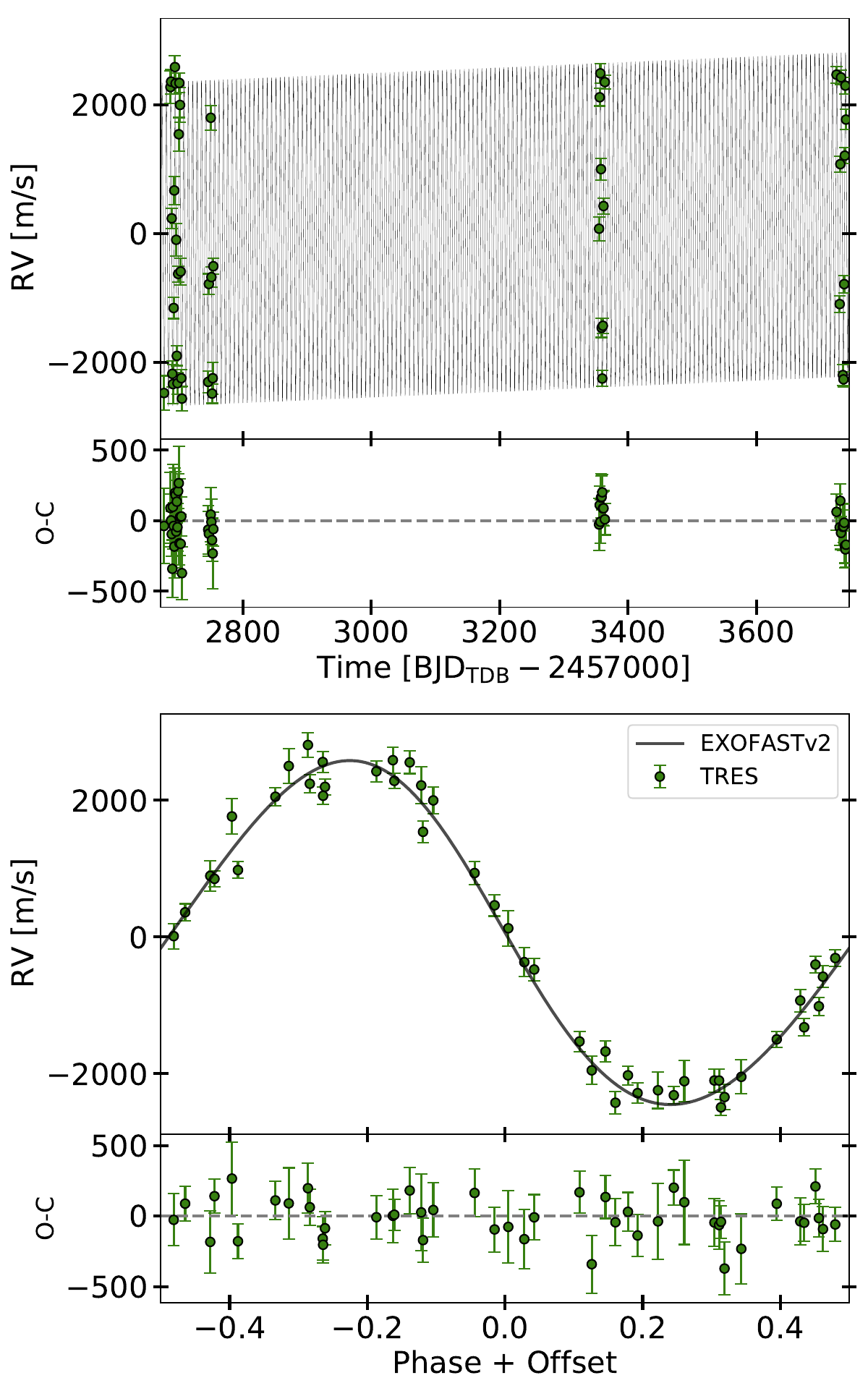}
    \caption{
    (Top): Best-fit radial velocity model showing full time domain of observations and a clear long-term velocity drift in the RVs.
    (Bottom): Relative multi-order RVs are plotted in green, phase-folded on the orbital period with the linear drift removed. The black line shows the best-fit RV curve from our analysis with \texttt{EXOFASTv2}. The orbital eccentricity is very close to zero (e = 0.054 $\pm$ 0.013).
    \label{fig:rv}}
\end{figure}

\begin{figure}[t!]
    \centering
    \includegraphics[width=\columnwidth]{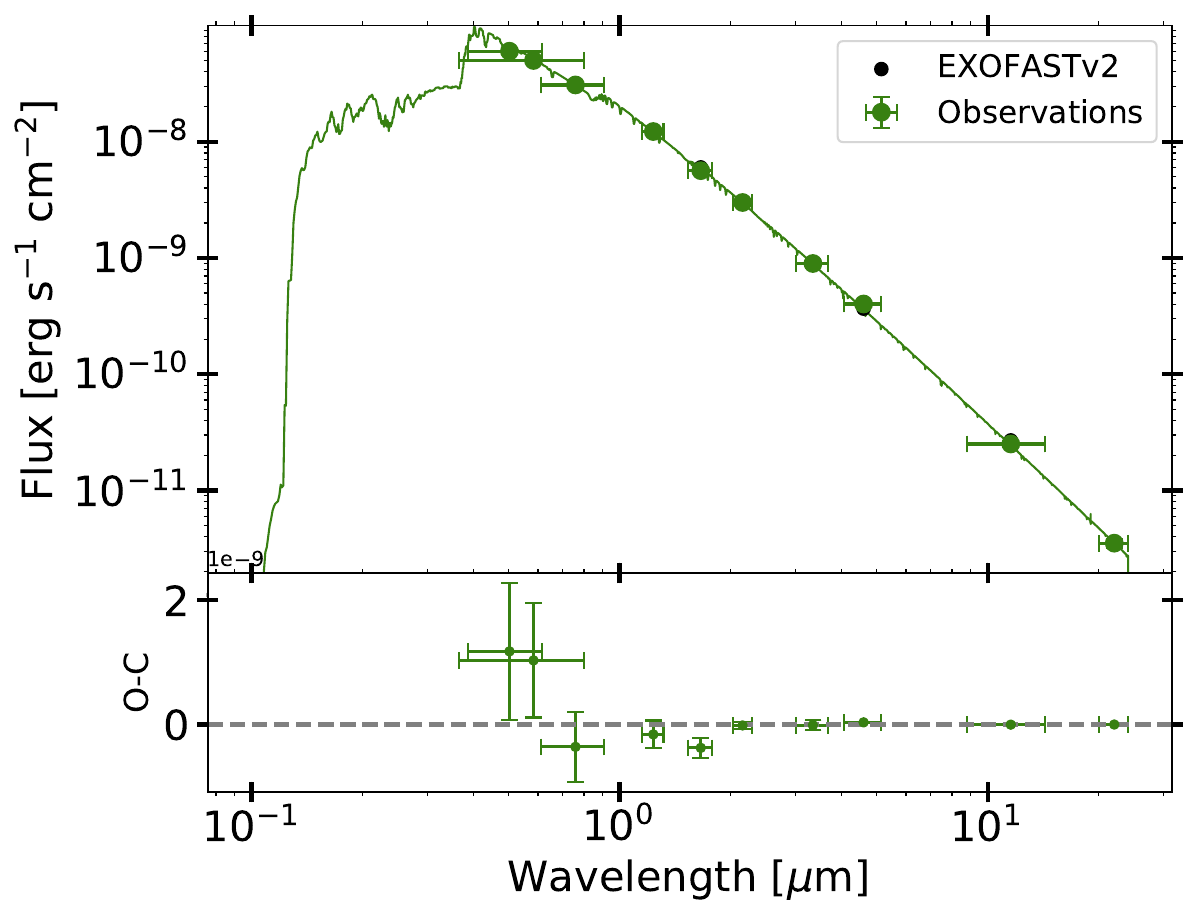}
    \caption{
    The spectral energy distribution of HIP 61637. Green points with error bars are the observed photometry, and the green line shows the corresponding best-fit stellar atmosphere model spectrum. Black points show the best-fit \texttt{EXOFASTv2} model flux integrated over each observed bandpass, used to constrain the stellar radius and effective temperature.
    \label{fig:sed}}
\end{figure}

\begin{deluxetable}{ccclDlc}[b]
\tablecaption{Multi-order relative RVs of HIP 61637, obtained with TRES.
\label{RV_table}}
\tablewidth{\columnwidth}
\tabletypesize{\scriptsize}
\tablehead{
\colhead{BJD\_UTC (days)} & 
\colhead{$v_{\text{rad}}$ (m s$^{-1}$)} &
\nocolhead{$h$} &
\colhead{$\sigma_{\text{RV}}$}} 
\startdata
2459676.783664 &    -4534.8 &&    258.5\\
2459686.779555 &      217.4 &&    242.8\\
2459687.811866 &      303.6 &&    175.7\\
2459688.820888 &    -1818.2 &&    143.2\\
2459689.784790 &    -4231.8 &&    192.9\\
2459690.703113 &    -4390.0 &&    291.7\\
2459691.853346 &    -3208.8 &&    150.4\\
2459692.829517 &    -1384.0 &&    209.9\\
2459693.798638 &      531.5 &&    163.3\\
2459694.808421 &      277.9 &&    149.6\\
2459695.785976 &    -2148.5 &&    248.3\\
2459696.750133 &    -3949.2 &&    136.1\\
2459697.830904 &    -4373.1 &&    153.8\\
2459698.833118 &    -2676.3 &&    103.6\\
2459699.873835 &     -508.5 &&    250.1\\
2459700.774790 &      288.2 &&    137.5\\
2459701.751607 &      -53.0 &&    262.3\\
2459702.774098 &    -2637.7 &&    198.5\\
2459703.802298 &    -4292.2 &&    119.4\\
2459704.760114 &    -4609.6 &&    174.6\\
2459745.681005 &    -4332.1 &&    157.6\\
2459746.711509 &    -2812.5 &&    143.5\\
2459749.674804 &     -230.9 &&    179.3\\
2459750.675842 &    -2703.8 &&    146.1\\
2459751.701421 &    -4511.0 &&    133.7\\
2459752.730332 &    -4272.1 &&    240.4\\
2459753.662326 &    -2533.7 &&     98.7\\
2460354.891002 &    -1689.1 &&    172.1\\
2460355.899358 &      351.4 &&    117.3\\
2460356.901693 &      723.2 &&    138.1\\
2460357.875529 &     -763.7 &&    155.0\\
2460358.916638 &    -3226.4 &&    132.7\\
2460359.852026 &    -4012.3 &&    105.3\\
2460360.873031 &    -3193.1 &&     97.3\\
2460361.833517 &    -1334.6 &&    103.6\\
2460363.908885 &      590.1 &&     87.2\\
2460725.013281 &      862.2 &&    108.5\\
2460729.917260 &    -2695.5 &&    110.0\\
2460730.896549 &     -523.2 &&     99.6\\
2460731.992961 &      823.4 &&     95.2\\
2460734.873765 &    -3795.2 &&    152.4\\
2460735.920990 &    -3861.5 &&     93.8\\
2460736.892110 &    -2382.9 &&    113.9\\
2460737.958806 &     -388.7 &&    101.7\\
2460738.803112 &      698.5 &&    111.0\\
2460739.794535 &      172.1 &&    142.2\\
\enddata
\vspace{-20pt} % Adjust the negative value as needed to remove extra space
%\tablecomments{\\$^a$ Barycentric Date of mid-exposure time, in the TBD time-system.\\$^b$ Barycentric RV.\\$^c$ Internal model-dependent uncertainties on each velocity are $\sigma$/$h$, where $\sigma$ is listed in Table 3 and $h$ is the peak-normalized cross-correlation for each spectrum listed here.\\$^d$ Peak-normalized cross-correlation.}
\end{deluxetable}

We obtained spectroscopic observations using TRES (\citealt{tres}) on the 1.5-m Tillinghast Reflector at the Fred Lawrence Whipple Observatory on Mount Hopkins, Arizona. TRES has a resolving power of R = 44,000 and covers a wavelength range of 390-910 nm with a typical spectral resolution element of $6.8 \kms$. The first two observations revealed broad lines corresponding to a projected equatorial velocity of about $80 \kms$, so we increased subsequent exposure times to a typical value of 460 seconds and signal-to-noise ratio per resolution element of 290.

We used multiple echelle orders to measure the RV of each observed spectrum, and derived precise relative velocities by cross-correlating individual observed spectra against a template constructed from all the observations. We then averaged the RVs from the orders that we decided were the most suitable. The full set of 46 multi-order relative RVs of HIP 61637 from TRES is given in Table \ref{RV_table}.

We derived the initial estimates for the host star effective temperature ($\teff=8681 \pm 751$\,K), metallicity ($\feh=-0.34 \pm 0.14$), and projected equatorial rotation velocity ($\vsini=79.2 \pm 1.6$ km s$^{-1}$) using the Stellar Parameter Classification tool (SPC) (\citealt{Buc12}). Due to the challenges of obtaining reliable SPC constraints for rapidly rotating A-type stars, we only use these parameters as starting values in our MCMC analysis, discussed in Section \ref{sec:analysis} along with the priors applied to the global fit.

We initially obtained 20 TRES observations at various phases of the orbit between April 7 and May 5, 2022. Our preliminary analysis showed a small long-term drift in the RVs of the system, prompting us to acquire additional observations. We obtained 7 more observations of HIP 61637 between June 15 and 23, 2022, an additional 9 observations between February 14 and 23, 2024, and a final set of 10 observations between February 18 and March 5, 2025, bringing the total to 46 TRES RVs spanning about three years. This extended baseline allowed us to constrain the long-term velocity drift to $0.426 \pm 0.054$ m s$^{-1}$day$^{-1}$, illustrated in Figure \ref{fig:rv}, and to account for the effect in our analysis.

A long-term RV drift can indicate the presence of an additional gravitationally bound object in the system. To explore this possibility, we first inspected nearby stars from the Gaia catalog but found no stellar neighbors with masses and separations consistent with the measured drift. This prompted us to examine existing high-resolution imaging data to search for a possible unresolved companion.

\subsection{High-resolution imaging}

The presence of a bound or unbound companion near a target star can lead to contamination of the observed light curve, affecting the transit depth and therefore the derived radius of the BD (\citealt{Cia15}). At the same time, such a companion can influence the radial velocities in the system, distorting the mass estimate of the transiting companion from RV measurements. Hence, it is important to examine the nearby field of a target star for possible contaminating stars at spatial resolutions tighter than the resolution of Gaia (\citealt{Car21}).

\begin{figure}[t]
    \centering
    \includegraphics[width=0.48\textwidth]{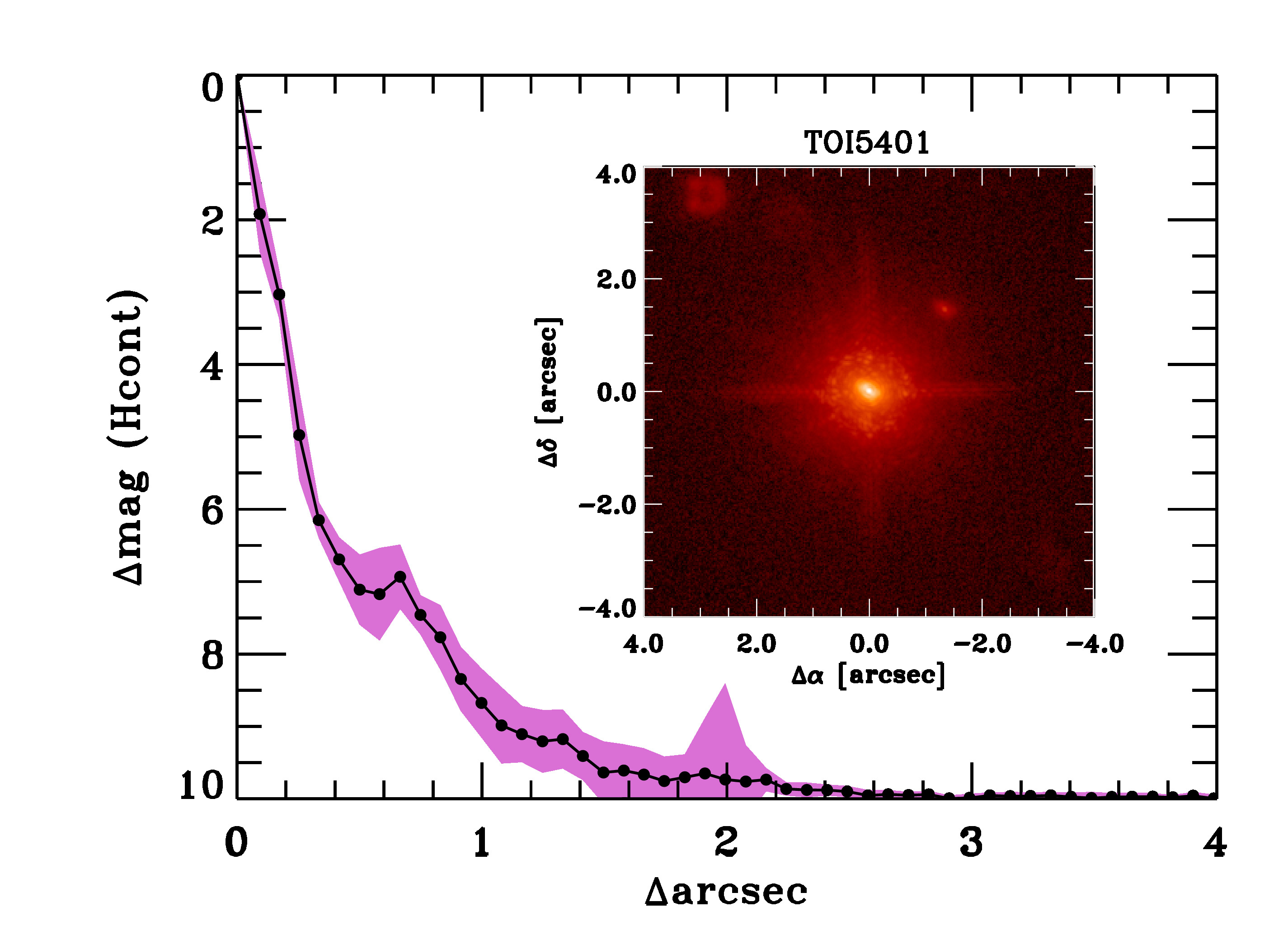}
    \caption{High-Resolution Imaging sensitivity curve obtained by \citet{Cia24} using the Palomar 5-meter telescope. The curve shows evidence of a nearby faint stellar companion within a 2-arcsecond separation from HIP 61637.
    \label{speckle_imaging}}
\end{figure}

We present the high-resolution imaging data and the corresponding sensitivity curve measured by \citet{Cia24} using the Palomar 5-m telescope in Figure \ref{speckle_imaging}. The imaging revealed a faint stellar companion within a 2-arcsecond separation, $317^\circ$ E of N from HIP 61637. The system was observed in two different (Kcont and Hcont) filters, in which the differences in magnitudes between HIP 61637 and its nearby companion were measured to be $\Delta \text{mag}_{K}$ = 5.841 $\pm$ 0.023 and $\Delta \text{mag}_{H}$ = 6.044 $\pm$ 0.026, respectively. Given that the infrared brightness ratios are more than 200, the impact of the stellar neighbor on the BD radius from TESS photometry is negligible.

The gravitational tug from a wide companion around a compact binary system would generally lead to a long-term velocity drift in the radial velocities. Given that we indeed measure a small but significant RV drift in the system, we checked whether this faint companion could be the source of the measured drift of $0.426 \pm 0.054$ m s$^{-1}$day$^{-1}$. Taking a lower bound for the separation between HIP 61637 and the faint companion, which corresponds to $420$ AU for a 2-arcsec separation subtended at a $212$ pc distance, and using mass-luminosity relations, we found that the maximum acceleration that could be induced by the companion is ~200 times smaller than the RV slope of HIP 61637. Therefore, the observed companion cannot be the source of the velocity drift. We therefore attribute this drift to another bound object that is too faint and too close to HIP 61637 to be resolved by our high-resolution imaging as documented by the sensitivity curve in Figure \ref{speckle_imaging}. 

\section{\textbf{Analysis}} \label{sec:analysis}

We used an MCMC analysis with \texttt{EXOFASTv2} (\citealt{Eas19}) to jointly model the radial velocity and transit photometry data, and determine the stellar and BD parameters. We made use of the evolved nature of HIP 61637 to obtain a reliable age estimate using stellar isochrones.

%\begin{figure}
%    \centering
%    \includegraphics[width=0.48\textwidth]{SED.png}
%    \caption{The spectral energy distribution of HIP 61637. The blue points are the predicted integrated fluxes for the star. The red error bars are observed values at the corresponding passbands. The horizontal error bars are the width of the passbands, and the vertical error bars represent the $1\sigma$ uncertainties. 
%    \label{fig4}}
%\end{figure}
\subsection{Global Model}

To measure the parameters of a transiting BD, it is essential to determine the stellar parameters of its host, as the BD mass and radius are relative to the host. We therefore performed a single, joint MCMC analysis with \texttt{EXOFASTv2}, simultaneously fitting the TESS transit photometry, our multi-order relative RVs from TRES, the Spectral Energy Distribution (SED) and parallax measurements from Gaia DR3, and MIST stellar isochrone models. The stellar parameters (e.g., mass, radius, and age) and the BD parameters were fit together within this single global model so that all of the above datasets and models could simultaneously inform the full set of stellar and BD parameters. We used a starting value of effective temperature from our SPC analysis of TRES spectra.

We placed Gaussian priors on two parameters in the global fit: the stellar metallicity ($\feh = 0 \pm 0.5$ dex), assuming a solar-neighborhood metallicity distribution for massive stars, and the Gaia DR3 parallax ($\varpi = 4.7195 \pm 0.0857$ mas). We also placed an upper bound on the line-of-sight extinction ($A_V < 0.08$) from Galactic dust maps (\citealt{Sch98}), and added a dilution term to the TESS band, with a Gaussian prior (0$\%$ $\pm$ 10$\%$) on the contamination ratio as reported by the TESS Input Catalogue (\citealt{Sta18}), to independently check that light from any nearby stars in the aperture was properly corrected. All other parameters were allowed to vary freely in the fit.

We considered the \texttt{EXOFASTv2} fit to be fully converged when it met the criteria described in \citealt{Eas19}, with Gelman-Rubin statistic less than 1.01 and independent draws greater than 1000.

%The spectral energy distribution of the star, which was used to determine initial values in the MCMC run, is shown in Figure \ref{fig4}. 
The best-fit transit and radial velocity models, along with the spectral energy distribution from the global \texttt{EXOFASTv2} fit, are shown in Figures \ref{fig:transits}, \ref{fig:rv}, and \ref{fig:sed}. The full set of stellar and BD parameters are reported in Table \ref{exofast_table}.

%The resulting posterior distributions for the stellar mass and age from the global MCMC analysis are presented in Figure \ref{posteriors}. 

\begin{figure}[t]
    \centering
    \includegraphics[width=0.48\textwidth]{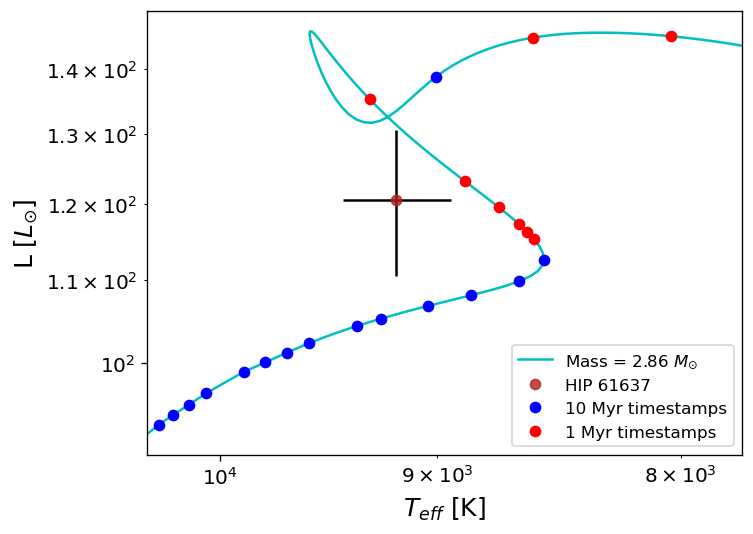}
    \caption{Luminosity-Temperature MIST evolutionary tracks with a constant mass of $2.86 \msun$. Blue and red circles mark equal timestamps of 50 and 1 Myr, respectively, along the evolutionary tracks. HIP 61637 is indicated with an orange circle.
    \label{evolutionary_track}}
\end{figure}

\subsection{Age Indicators for HIP 61637}
\label{sec:age_indicators}

Our global fit of the system revealed that HIP 61637 is nearing the end of its lifetime on the main sequence and has begun to evolve, making it a good candidate for a reliable age determination with stellar evolutionary tracks. We used this opportunity to model the stellar age using MIST evolutionary tracks (\citealt{Dot16}, \citealt{Cho16}, \citealt{Pax11, Pax13, Pax15}) built into \texttt{EXOFASTv2} and found an age of 396 $\pm$ 46 Myr. In Figure \ref{evolutionary_track}, we show the MIST stellar evolutionary track with a constant mass of $2.86 M_{\odot}$ overplotted on the parameters determined for HIP 61637 b. The blue and red circles along the evolutionary tracks indicate equal timestamps of 50 and 1 Myr, respectively, and highlight that the star is evolving fast enough in the luminosity-temperature parameter space to allow for a tight constraint on the inferred age.

\begin{figure}[b]
    \centering
    \includegraphics[width=0.47\textwidth]{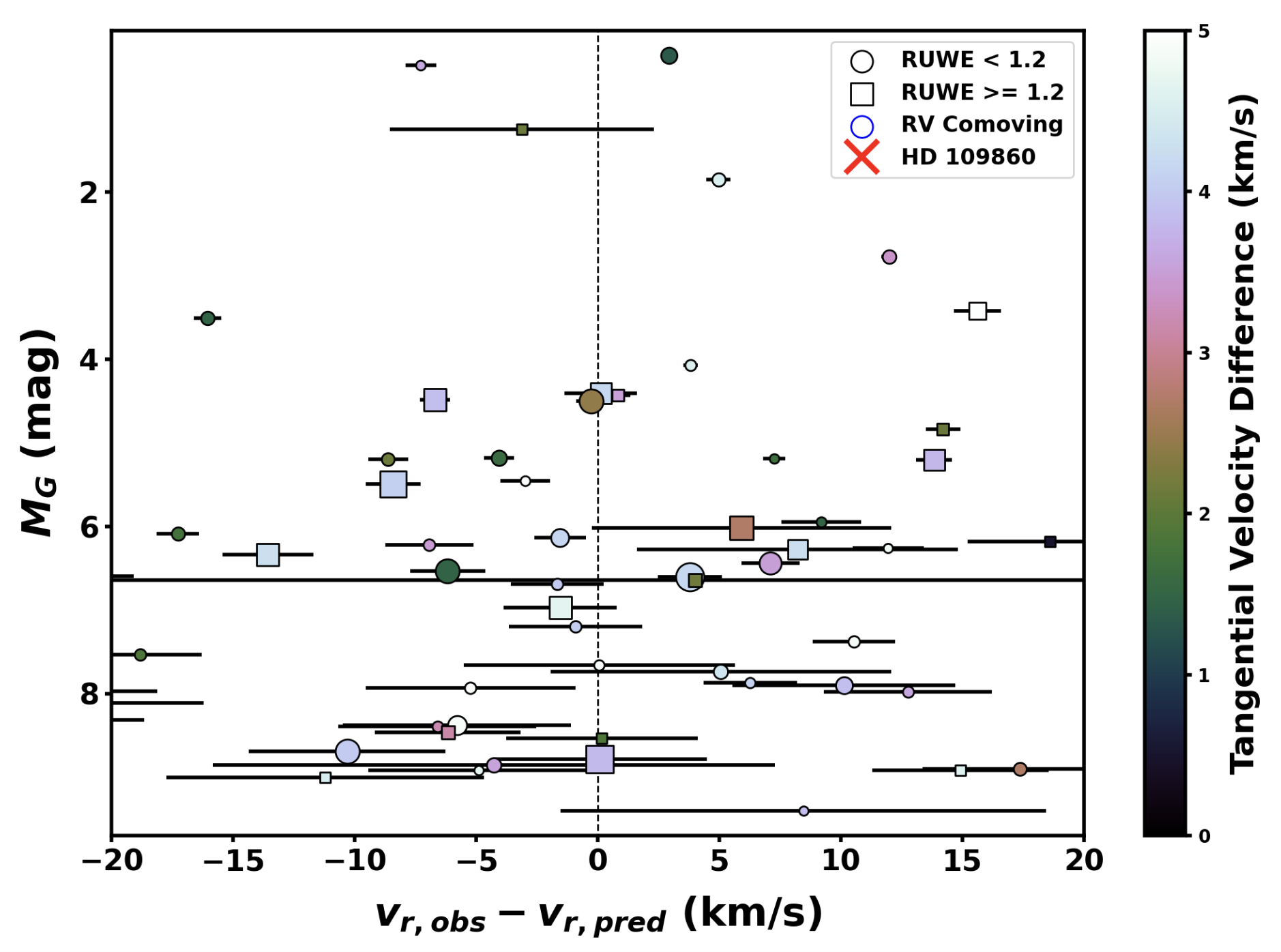}
    \caption{Magnitude-RV-offset distribution of all stars within 50 pc separation and within 20 km s$^{-1}$ radial velocity offset from HIP 61637. The symbol sizes represent 3D distance from HIP 61637. The random distribution shows no sign of a comoving group.
    \label{no_cluster}}
\end{figure}

Given the relatively young age of the host, we analyzed the surrounding field of stars using the \texttt{FriendFinder} (\citealt{FriendFinder}) tool to test for the membership of HIP 61637 in a young stellar association. We plot all co-moving stars in the neighborhood with a 20 km s$^{-1}$ radial velocity offset from HIP 61637 in Figure \ref{no_cluster}. The distribution in the velocity offsets in the neighboring stars appears random, and there is no sign of a narrow range of velocities centered around that of HIP 61637. We conclude that there is no evidence for a membership in a comoving cluster.

It is difficult to find other independent age indicators to constrain the host's age. We found no Li I 6708 Å absorption in the stellar spectra. However, the rapid rotation of HIP 61637 makes it unclear whether lithium is truly absent from its surface, or is simply masked by rotational broadening of the spectra, making it difficult to place a lower bound on the age of the host star. Similarly, no Ca II H\&K or Ca II infrared triplet emission lines were detected in our spectra of HIP 61637. Finally, gyrochronology is not well-suited for the host star, as estimates are less reliable for rapid rotators, and period-age relationships have not been well-studied for A-type stars.

\subsection{Circularization Timescales}
\label{sec:tidal_circ}

Given our constraint on the age of HIP 61637 b from stellar isochrones, we explore the role that tidal interactions with its host star could have played in the evolution of its orbit and rotation. To this end, we examine the predicted orbital circularization timescales using the stellar and companion properties derived with EXOFASTv2. We use the equation for orbital circularization for close-in extrasolar planets given by \cite{Jac08}. The circularization timescale, $\tau_\text{e}$, of the system is
\begin{equation}
\label{eq:tau1}
    \frac{1}{\tau_{\text{circ}, *}} = \frac{171}{16} \sqrt{\frac{G}{M_*}} \frac{R_*^5 M_{\text{BD}}}{Q_*} a^{ -\frac{13}{2} }
\end{equation}

\begin{equation}
\label{eq:tau2}
    \frac{1}{\tau_{\text{circ},\text{BD}}} = \frac{63}{4} \frac{ \sqrt{GM_*^3} {R_{\text{BD}}}^5} {Q_{\text{BD}} M_{\text{BD}}} a^{ -\frac{13}{2} }
\end{equation}

\begin{equation}
\label{eq:tau3}
    \frac{1}{\tau_\text{e}} = \frac{1}{\tau_{\text{circ}, *}} + \frac{1}{\tau_{\text{circ},\text{BD}}},
\end{equation}

where $a$ is the semi-major axis, $M_*$ is the stellar mass, $R_*$ is the stellar radius, $M_{\mathrm{BD}}$ is the BD mass, $R_{\mathrm{BD}}$ is the BD radius, and $Q_*$ and $Q_{\mathrm{BD}}$ are the tidal quality factors for the star and the BD. The equations (\ref{eq:tau1}) to (\ref{eq:tau3}) assume that $e$ values are small and describe orbit-averaged effects of the tides. The stellar tide $Q_*$ describes the tide raised on the star by the BD, and $Q_{\mathrm{BD}}$ reflects the tide raised on the BD by the star. These tidal factors are similar to the modified quality factor $Q'$ in \cite{Gol66}.

\begin{figure}
    \centering
    \includegraphics[width=0.46\textwidth]{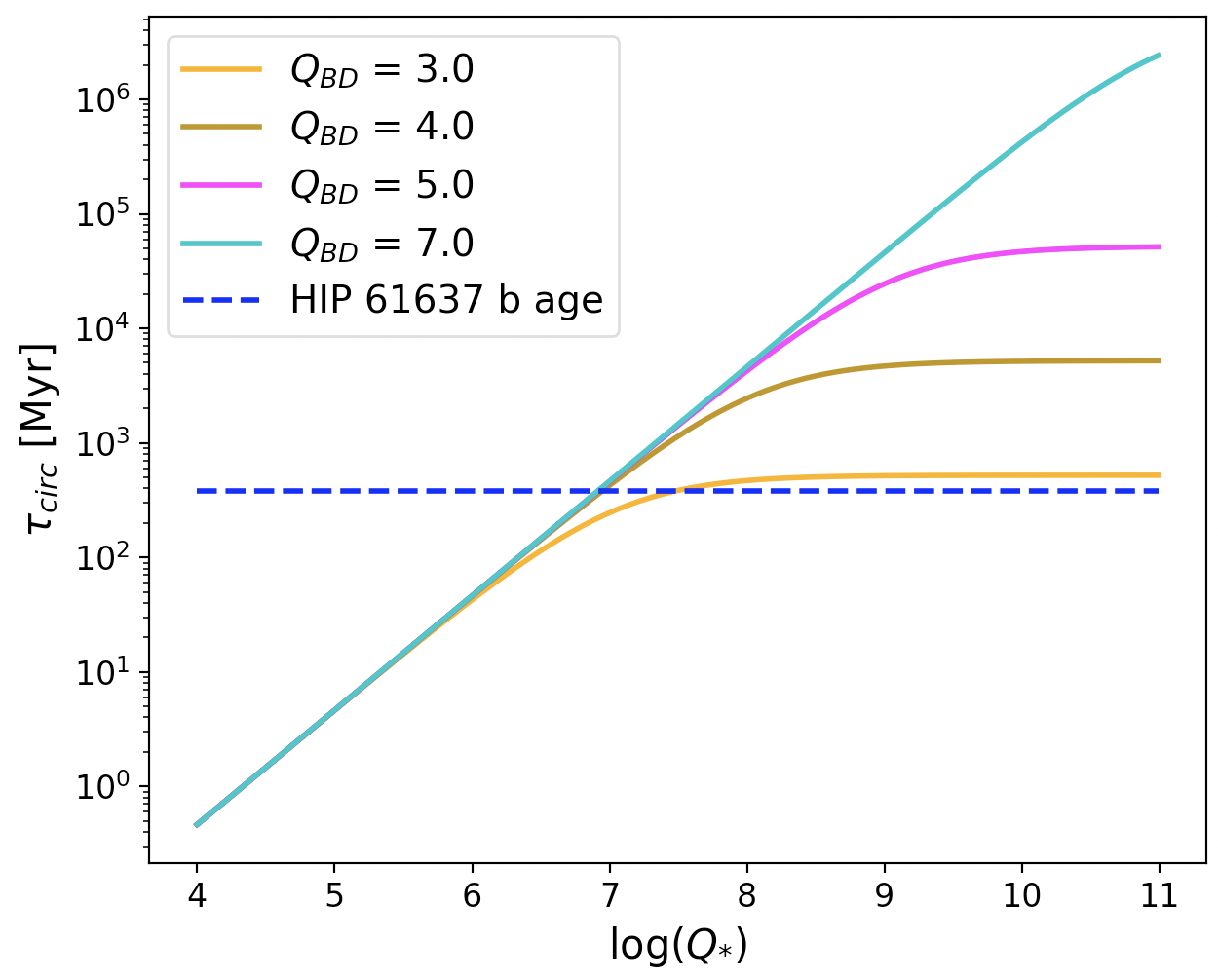}
    \caption{Tidal Circularization timescales for possible ranges of stellar and BD quality factors. The timescale is plotted as a function of $Q_*$, with the four different lines showing $Q_{\mathrm{BD}}$ values of $10^{3}$, $10^{4}$, $10^5$, and $10^6$.
    \label{tau_circ}}
\end{figure}

The tidal quality factors of stars and BDs are difficult to constrain (\citealt{Sub20}). Therefore, we calculate the tidal circularization timescales for a broad range of stellar ($Q_* = 10^4 - 10^{11}$) and BD ($Q_{\text{BD}} = 10^3 - 10^7$) tidal quality factors, shown in Figure \ref{tau_circ}. We find that stellar damping plays the dominant role in the circularization of the system until $Q_* = ~10^8$, after which point the $Q_{\mathrm{BD}}$ starts to influence the circularization timescale. The age of the system exceeds $\tau_{\mathrm{circ}}$ if $Q_*<10^7$. In other words, if the tidal dissipation mechanisms are as effective as $Q_*<10^7$ in the star or $Q_{\mathrm{BD}} < 10^3$ in the BD, the orbit of HIP 61637 b would have been subject to circularization.

Recent work (\citealt{Ess24}) has made progress on the theoretical understanding of tidal dissipation mechanisms throughout the evolutionary stages of intermediate-mass stars. We use this opportunity to estimate the $Q_*$ of HIP 61637. The modified tidal quality factor $Q'$, which is similar to $Q_*$ in \cite{Jac08}, is related to the Love number $k$ by:
\begin{equation}
    Q' = \frac{2 |k(\omega)|}{3 |\text{Im}(k(\omega))| \mathrm{Re}(k(\omega))} \approx \frac{2}{3 |\text{Im}(k(\omega))|},
\end{equation}
where for a weakly dissipative stellar fluid $|k| \approx \mathrm{Re}(k)$ (\citealt{Ess24}). The authors model intermediate-mass stars ($1-4\,M_\odot$) from the main sequence through the giant branches. For their $3.5\,M_\odot$ track with a companion in a 7-day orbit (their $2\,M_\odot$ and $3.5\,M_\odot$ models bracket HIP 61637, $M_* = 2.86\,M_\odot$), the Love numbers remain roughly constant over the majority of the main-sequence lifetime, with $|\mathrm{Im}(k)_{\mathrm{eq}}| = 10^{-6}$ and $|\mathrm{Im}(k)_{\mathrm{IGW}}| = 10^{-2}$. Although the equilibrium tide carries the larger quasi-static deformation, the dynamical tide (internal gravity waves excited at the convective-core boundary) provides the dominant dissipation at these short orbital periods.

As the star crosses the Hertzsprung gap, the convective core vanishes and the tidal dissipation passes through a minimum. The strong enhancement of the equilibrium tide by a deep convective envelope sets in only later, on the red giant branch. With $T_{\rm eff} = 9180$\,K and $R_* = 4.33\,R_\odot$, HIP 61637 is still at the main-sequence turnoff and retains a radiative envelope, so this enhancement has not yet occurred. Because the circularization is integrated over the long main-sequence phase, we adopt the main-sequence dissipation: the dominant dynamical tide ($|\mathrm{Im}(k)| \approx 10^{-2}$) gives a time-averaged $\langle Q_* \rangle \sim 10^2$, while even the subdominant equilibrium tide alone ($|\mathrm{Im}(k)| \approx 10^{-6}$) gives $\langle Q_* \rangle \sim 10^6$. In either case $Q_* < 10^7$, so this theory of tidal dissipation indicates that the orbit of HIP 61637 b has been subject to tidal circularization.

\section{\textbf{Discussion}}

The newly discovered BD, HIP 61637 b, contributes to the small but growing sample of well-characterized transiting BDs from the TESS mission. The evolutionary stage of the host star allows for tight age estimation of the system, making HIP 61637 b a new addition to only 7 known transiting BDs with well-determined ages (\citealt{Gil17}, \citealt{Dav19}, \citealt{Car19}, \citealt{Car20, Car21}, \citealt{Vow23}). This gives us the opportunity to observationally test both recent substellar evolutionary models in a sparsely populated BD region and theories of tidal dissipation in lesser-studied intermediate-mass stars.

%\subsection{HIP 61637 b Properties}
%With a mass of $M_{\mathrm{BD}}$ = 46.2 $\pm$ 2.4 $M_J$, HIP 61637 b lies right at the middle of the traditional 13-80 $M_J$ mass limits that define a BD. The transiting BD has an orbital period of $P$ = 6.829103 $\pm$ 0.000011 days, and a near-eccentric orbit with $e$ = 0.071 $\pm$ 0.021. Its host star is an A-type sub-giant, with a mass of $M_*$ = 2.68 $\pm$ 0.20 \msun, radius of $R_*$ = 4.36 $\pm$ 0.16 \rsun, effective temperature of $T_{\text{eff}}$ = 9260 $\pm$ 270 K, surface gravity of log $g$ = 3.590 $\pm$ 0.044 cgs, and metallicity of [Fe/H] = -0.12 $\pm$ 0.21 dex. A full list of the stellar and BD parameters from \texttt{EXOFASTv2} is given in Table \textcolor{blue}{2}.

\begin{figure*}[ht]
\gridline{\fig{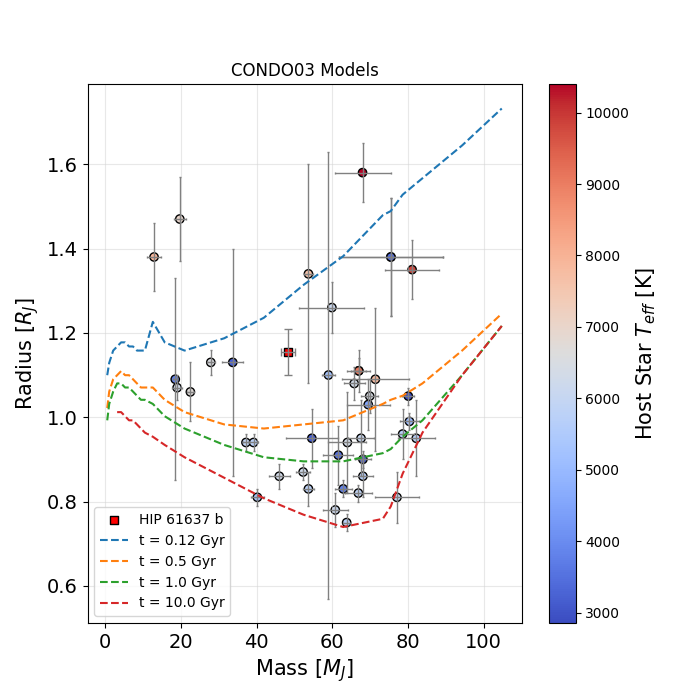}{0.5\textwidth}{(a)}
          \fig{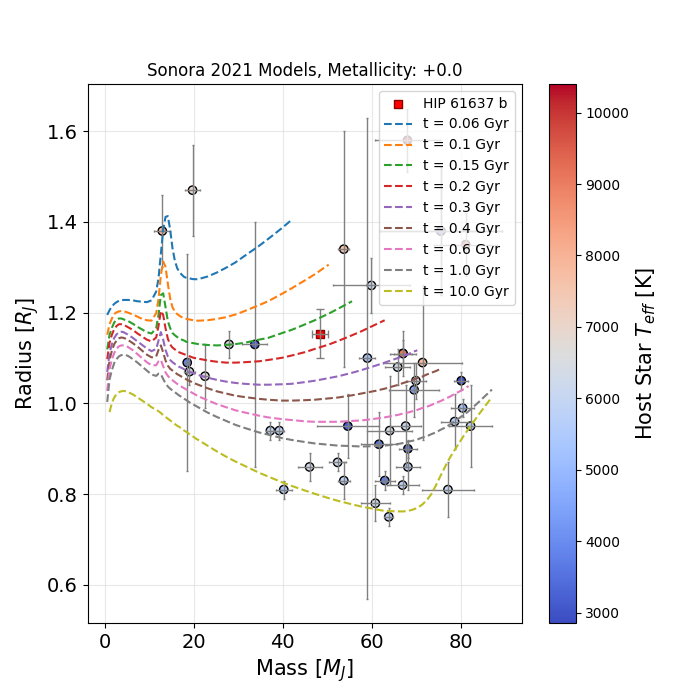}{0.5\textwidth}{(b)}}
\caption{Mass-radius diagram of all known transiting BDs \citep{Carmichael2023}. The (a) COND03 and (b) Sonora (2021) substellar evolutionary models are overplotted for different ages of BDs. HIP 61637 is shown with a red square.
\label{ev_models}}
\end{figure*}

\subsection{Testing Substellar Evolution Models}

Our precise characterization of HIP 61637 b, including the age determination of the system, allows us to test substellar evolution models in the sparsely populated region of the BD desert. We assume that the host star and its companion BD formed at the same time, thereby taking the age of HIP 61637 from stellar isochrones as the age of the BD. Given that the evolution of stars is significantly better understood than that of BDs, we take the approach of testing substellar evolution models while relying on ages from stellar isochrones.

We consult Sonora models by \cite{Mar21} and COND03 models by \cite{Bar03} to examine how well they predict the age of HIP 61637 b based on our observed parameters. COND03 models present isochrones for giant exoplanets and cool BDs that take into account irradiation from the host star. Sonora 2021 models do not incorporate irradiation, but do account for different host star metallicities.

We show all known transiting BDs on a radius-mass diagram in Figure \ref{ev_models}, together with lines of constant age from the COND03 and Sonora (2021) models. HIP 61637 b occupies a region in the radius-mass diagram of the transiting BD population that is not well-populated, making it an important contribution toward testing such evolutionary models. We find that the COND03 model constrains the age of HIP 61637 b between 120 and 500 Myr, in rough agreement with our independent age estimate of HIP 61637 b.

The Sonora (2021) evolutionary models place tighter constraints on the age, between 150 and 200 Myrs, which is inconsistent with our age determination. This may be explained by the fact that these models do not consider the effects of stellar irradiation. At a distance of only 0.1 AU from its A-type host star, irradiation effects will be important in the evolution of HIP 61637 b, inflating the radius of the BD. This means that HIP 61637 b will take longer to cool down and contract in size and, therefore, appear to be younger compared to its true age than in the scenario considered by the models.

The tension between the ages predicted by the well-studied stellar isochrones of A-type stars and the much lesser-known evolutionary pathways of BDs indicates a need for a more complete set of substellar models that incorporate wider considerations, particularly the effects of irradiation from the host star on BDs. Nevertheless, it is interesting to consider whether inflation in BD size from irradiation alone is enough to explain these inconsistencies, or whether another mechanism that current models do not capture might be at play. This issue is beyond the scope of this paper.

\subsection{HIP 61637 b in the Context of the Transiting BD Population}
With a mass of $M_*$ = 2.86 $\pm$ 0.12 \msun, HIP 61637 is the brightest and the most massive star to date transited by a BD companion. With a surface temperature of $T_{\text{eff}}$ = 9180 $\pm$ 240 K, it is also the third hottest, and one of only 6 A-type stars with such companions (\citealt{Zho19}, \citealt{Sub20}, \citealt{Gri21}, \citealt{Psa22}, \citealt{Seb22}, \citealt{Vow23}). A full list of the stellar and BD parameters from \texttt{EXOFASTv2} is given in Table \ref{exofast_table}.

Among the known transiting BD systems, HIP 61637 presents itself as an interesting counterpoint to the recently discovered HIP 33609 (TOI-588) system (\citealt{Vow23}). We present a comparison of key parameters for the two systems in Table \ref{tab:brown_dwarfs}. The two BDs have similar masses and both orbit hot, massive, and rapidly rotating A and B-type stars. The systems are $\sim150$ and $\sim400$ Myr in age, and HIP 61637 is enough older to have progressed further along the main sequence and has started to evolve. The main difference between the two systems lies in the orbital eccentricities and periods. While the $6.8$-day orbit of HIP 61637 b is nearly circular, the $39.4$-day orbit of HIP 33609 b is highly eccentric ($e = 0.56 \pm 0.03$). According to the tidal circularization theory presented in Section \ref{sec:tidal_circ}, HIP 33609 b is virtually immune to circularization effects over its short lifetime of $150$ Myr. Thus its eccentric orbit may be primordial, preserved since its formation, consistent with the coplanar high-eccentricity migration favored for HIP 33609 b based on its low measured obliquity (\citealt{Vow26}).  On the other hand, in Section \ref{sec:tidal_circ} we concluded that HIP 61637 b has been subjected to tidal circularization, due to its much tighter orbit. This is consistent with its near-zero but statistically significant eccentricity, as the system heads asymptotically towards circular.

\begin{table}[h!]
\centering
\scriptsize
\renewcommand{\arraystretch}{1.1}
\caption{Parameter comparison and circularization status of HIP 61637 and HIP 33609 systems}

\begin{tabular*}{\linewidth}{@{\extracolsep{\fill}}lcc}
\hline\hline
\rule{0pt}{3.2ex}\textbf{Parameter} & \shortstack{\rule{0pt}{3.2ex}\textbf{HIP 61637}\\[3pt](TOI-5401)} & \shortstack{\rule{0pt}{3.2ex}\textbf{HIP 33609}\\[3pt](TOI-588)} \rule[-2ex]{0pt}{0pt}\\
\hline
BD Mass ($M_{\mathrm{Jup}}$) & $47.8 \pm 1.5$ & $68.0 \pm 7.4$ \\
Stellar Mass ($M_\odot$) & $2.86 \pm 0.12$ & $2.38 \pm 0.10$ \\
Temperature (K) & $9,180 \pm 240$ & $10,400 \pm 800$ \\
Period (days) & $6.829$ & $39.372$ \\
Age (Myr) & $396 \pm 46$ & $153 \pm 24$ \\
$v\sin{i_*}$ (km s$^{-1}$) & $79.2 \pm 1.6$ & $55.6 \pm 1.8$ \\
Eccentricity & $0.054 \pm 0.013$ & $0.56 \pm 0.03$ \\
\hline
\end{tabular*}

\label{tab:brown_dwarfs}
\end{table}

\subsection{How did HIP 61637 b form?}

Astronomers have long debated about how BDs form. 
The traditional definition of BDs was arbitrarily based on the physical conditions needed for the fusion of hydrogen and deuterium. By these standards, a BD is an object with a mass between 13 and 80 Jupiter masses, making it massive enough to fuse deuterium, but not massive enough to sustain the temperatures and pressures needed for hydrogen fusion. However, these arbitrary mass limits do not account for the formation processes of BDs, and two different objects that fall within this same mass range, and are thereby labeled ``brown dwarfs'', could have formed by completely different mechanisms.

\citet{Car20} argue that BDs span the space occupied by the tail ends of two mass distributions: typically lower-mass companions that form like giant planets by core accretion, and generally higher-mass companions that form like low-mass binary stars by fragmentation of a molecular cloud. HIP 61637 b lies right in the middle of the ``brown dwarf desert" where these two distributions may overlap. As the short period of the system subjects HIP 61637 b to tidal circularization, the orbital eccentricity alone does not distinguish between the two mechanisms. The BD may have formed like a companion star with an eccentric orbit that has been nearly fully circularized, or like a planet by accretion in a disk with a nearly circular orbit that has been preserved. While forming a 50-Jupiter mass object by core accretion is challenging, the high mass of the HIP 61637 system may have provided unusual conditions for the formation of an exceptionally massive planet. Both explanations stretch the limits of our theoretical understanding of the physics involved in BD formation. In the future, with population-level catalogs from TESS and atmospheric studies of BDs from JWST and Ariel, we can hope to more carefully probe the BD desert and disentangle the different mechanisms responsible for forming these objects.

\subsection{Opportunities for Follow-up Observations}
HIP 61637 b is well suited for follow-up observations. The system has a short orbital period of $P$ = 6.829104 $\pm$ 0.000011 days that offers many opportunities to observe transits, although the long transit duration of $\sim 10$ hours poses a challenge for observing full transit events from the ground. Despite its rapid rotation of $v\sin{i} \approx 80$ km s$^{-1}$, the high SNR enabled by the remarkable brightness of the star facilitates high quality spectroscopic observations. Unfortunately, HIP 61637 is too bright for follow-up observations with JWST. Regardless, transit spectroscopy of atmospheres of intermediate-mass BDs such as HIP 61637 b, with facilities such as JWST and Ariel, may allow us to compare stellar and BD compositions and provide us with important clues about BD formation mechanisms.

\section{\textbf{Summary}}
We present the discovery and characterization of a new BD from NASA's TESS mission, in the middle of the $13 - 80$ $M_J$ mass limits that have traditionally defined BDs. HIP 61637 b was discovered in the primary mission using the 30 minute cadence, full-frame images, and orbits a roughly 400-Myr-old sub-giant A-type star in a nearly circular orbit, with a period of 6.8 days. This discovery adds to a handful of well-characterized transiting BDs with reliable age estimates, allowing us to test evolutionary models for substellar companions. At the same time, HIP 61637 is the most massive star known to host a BD, creating interesting possibilities for the formation scenarios of the system. Although TESS has contributed greatly to our understanding of BDs, increasing the number of transiting BDs will continue to advance our understanding of their formation and evolution.

% \texttt{EXOFASTv2} Results Table
\startlongtable
\begin{deluxetable*}{lcc}
\tablecaption{Median values and 68\% confidence interval for HIP 61637, created using EXOFASTv2}
\tablehead{\colhead{~~~Parameter} & \colhead{Units} & \colhead{Values}}
\startdata
\smallskip\\\multicolumn{2}{l}{Stellar Parameters:}&\smallskip\\
~~~~$M_*$\dotfill &Mass (\msun)\dotfill &$2.86^{+0.12}_{-0.11}$\\
~~~~$R_*$\dotfill &Radius (\rsun)\dotfill &$4.33^{+0.17}_{-0.13}$\\
% ~~~~$R_{*,SED}$\dotfill &Radius$^{1}$ (\rsun)\dotfill &$4.35^{+0.11}_{-0.10}$\\
~~~~$L_*$\dotfill &Luminosity (\lsun)\dotfill &$120.5^{+10}_{-9.1}$\\
%~~~~$F_{Bol}$\dotfill &Bolometric Flux (cgs)\dotfill &$0.0000000856^{+0.0000000067}_{-0.0000000057}$\\
~~~~$\rho_*$\dotfill &Density (cgs)\dotfill &$0.0499^{+0.0047}_{-0.0053}$\\
~~~~$\log{g}$\dotfill &Surface gravity (cgs)\dotfill &$3.622^{+0.028}_{-0.033}$\\
~~~~$T_{\rm eff}$\dotfill &Effective Temperature (K)\dotfill &$9180^{+240}_{-230}$\\
% ~~~~$T_{\rm eff,SED}$\dotfill &Effective Temperature$^{1}$ (K)\dotfill &$9170^{+230}_{-200}$\\
~~~~$[{\rm Fe/H}]$\dotfill &Metallicity (dex)\dotfill &$0.04^{+0.12}_{-0.11}$\\
~~~~$[{\rm Fe/H}]_{0}$\dotfill &Initial Metallicity$^{2}$ \dotfill &$0.03^{+0.12}_{-0.10}$\\
~~~~$Age$\dotfill &Age (Gyr)\dotfill &$0.396^{+0.046}_{-0.042}$\\
~~~~$EEP$\dotfill &Equal Evolutionary Phase$^{3}$ \dotfill &$397.0^{+5.0}_{-5.3}$\\
~~~~$A_V$\dotfill &V-band extinction (mag)\dotfill &$0.041^{+0.024}_{-0.027}$\\
~~~~$\sigma_{SED}$\dotfill &SED photometry error scaling \dotfill &$1.90^{+0.66}_{-0.42}$\\
~~~~$\varpi$\dotfill &Parallax (mas)\dotfill &$4.715^{+0.083}_{-0.084}$\\
~~~~$d$\dotfill &Distance (pc)\dotfill &$212.1^{+3.8}_{-3.7}$\\
~~~~$\dot{\gamma}$\dotfill &RV slope$^{4}$ (m s$^{-1}$day$^{-1}$)\dotfill &$0.426^{+0.054}_{-0.053}$\\
\smallskip\\\multicolumn{2}{l}{Planetary Parameters:}&b\smallskip\\
~~~~$P$\dotfill &Period (days)\dotfill &$6.829104\pm0.000011$\\
~~~~$R_P$\dotfill &Radius (\rj)\dotfill &$1.149^{+0.049}_{-0.038}$\\
~~~~$M_P$\dotfill &Mass (\mj)\dotfill &$47.8^{+1.5}_{-1.4}$\\
~~~~$T_C$\dotfill &Time of conjunction$^{5}$ (\bjdtdb)\dotfill &$2458930.89420^{+0.00100}_{-0.00095}$\\
~~~~$T_T$\dotfill &Time of minimum projected separation$^{6}$ (\bjdtdb)\dotfill &$2458930.89366^{+0.00100}_{-0.00095}$\\
~~~~$T_0$\dotfill &Optimal conjunction Time$^{7}$ (\bjdtdb)\dotfill &$2459463.56421^{+0.00049}_{-0.00048}$\\
~~~~$a$\dotfill &Semi-major axis (AU)\dotfill &$0.1005^{+0.0014}_{-0.0013}$\\
~~~~$i$\dotfill &Inclination (Degrees)\dotfill &$86.6^{+1.9}_{-1.5}$\\
~~~~$e$\dotfill &Eccentricity \dotfill &$0.054\pm0.013$\\
~~~~$\omega_*$\dotfill &Argument of Periastron (Degrees)\dotfill &$64.9^{+7.2}_{-9.3}$\\
~~~~$T_{eq}$\dotfill &Equilibrium temperature$^{8}$ (K)\dotfill &$2908^{+60}_{-54}$\\
~~~~$\tau_{\rm circ}$\dotfill &Tidal circularization timescale (Gyr)\dotfill &$600\pm110$\\
~~~~$K$\dotfill &RV semi-amplitude (m/s)\dotfill &$2512^{+34}_{-32}$\\
~~~~$R_P/R_*$\dotfill &Radius of planet in stellar radii \dotfill &$0.02730^{+0.00016}_{-0.00015}$\\
~~~~$a/R_*$\dotfill &Semi-major axis in stellar radii \dotfill &$5.00^{+0.15}_{-0.18}$\\
~~~~$\delta$\dotfill &$\left(R_P/R_*\right)^2$ \dotfill &$0.0007454^{+0.0000088}_{-0.0000083}$\\
~~~~$\delta_{\rm TESS}$\dotfill &Transit depth in TESS (fraction)\dotfill &$0.0008046^{+0.0000087}_{-0.0000086}$\\
~~~~$\tau$\dotfill &Ingress/egress transit duration (days)\dotfill &$0.01201^{+0.0011}_{-0.00081}$\\
~~~~$T_{14}$\dotfill &Total transit duration (days)\dotfill &$0.4127^{+0.0015}_{-0.0013}$\\
~~~~$T_{FWHM}$\dotfill &FWHM transit duration (days)\dotfill &$0.4006\pm0.0010$\\
~~~~$b$\dotfill &Transit Impact parameter \dotfill &$0.28^{+0.11}_{-0.16}$\\
~~~~$b_S$\dotfill &Eclipse impact parameter \dotfill &$0.31^{+0.12}_{-0.17}$\\
~~~~$\tau_S$\dotfill &Ingress/egress eclipse duration (days)\dotfill &$0.01336^{+0.0013}_{-0.00093}$\\
~~~~$T_{S,14}$\dotfill &Total eclipse duration (days)\dotfill &$0.450^{+0.012}_{-0.011}$\\
~~~~$T_{S,FWHM}$\dotfill &FWHM eclipse duration (days)\dotfill &$0.436^{+0.012}_{-0.011}$\\
~~~~$\delta_{S,2.5\mu m}$\dotfill &Blackbody eclipse depth at 2.5$\mu$m (ppm)\dotfill &$104.1^{+3.9}_{-3.4}$\\
~~~~$\delta_{S,5.0\mu m}$\dotfill &Blackbody eclipse depth at 5.0$\mu$m (ppm)\dotfill &$162.1^{+5.1}_{-4.2}$\\
~~~~$\delta_{S,7.5\mu m}$\dotfill &Blackbody eclipse depth at 7.5$\mu$m (ppm)\dotfill &$185.0^{+5.6}_{-4.5}$\\
~~~~$\rho_P$\dotfill &Density (cgs)\dotfill &$39.1^{+4.0}_{-4.5}$\\
~~~~$logg_P$\dotfill &Surface gravity \dotfill &$4.953^{+0.029}_{-0.036}$\\
~~~~$\Theta$\dotfill &Safronov Number \dotfill &$2.92^{+0.11}_{-0.12}$\\
~~~~$\fave$\dotfill &Incident Flux (\fluxcgs)\dotfill &$16.2^{+1.4}_{-1.2}$\\
~~~~$T_P$\dotfill &Time of Periastron (\bjdtdb)\dotfill &$2458930.47^{+0.13}_{-0.17}$\\
~~~~$T_S$\dotfill &Time of eclipse (\bjdtdb)\dotfill &$2458927.577\pm0.026$\\
~~~~$T_A$\dotfill &Time of Ascending Node (\bjdtdb)\dotfill &$2458936.166^{+0.031}_{-0.033}$\\
~~~~$T_D$\dotfill &Time of Descending Node (\bjdtdb)\dotfill &$2458932.542^{+0.032}_{-0.031}$\\
~~~~$V_c/V_e$\dotfill & \dotfill &$0.953\pm0.013$\\
~~~~$e\cos{\omega_*}$\dotfill & \dotfill &$0.0225^{+0.0061}_{-0.0060}$\\
~~~~$e\sin{\omega_*}$\dotfill & \dotfill &$0.048^{+0.013}_{-0.014}$\\
% ~~~~$M_P\sin i$\dotfill &Minimum mass (\mj)\dotfill &$47.7^{+1.5}_{-1.4}$\\
% ~~~~$M_P/M_*$\dotfill &Mass ratio \dotfill &$0.01595\pm0.00031$\\
% ~~~~$d/R_*$\dotfill &Separation at mid transit \dotfill &$4.75\pm0.18$\\
% ~~~~$P_T$\dotfill &A priori non-grazing transit prob \dotfill &$0.2049^{+0.0083}_{-0.0073}$\\
% ~~~~$P_{T,G}$\dotfill &A priori transit prob \dotfill &$0.2165^{+0.0088}_{-0.0077}$\\
% ~~~~$P_S$\dotfill &A priori non-grazing eclipse prob \dotfill &$0.1856^{+0.0074}_{-0.0058}$\\
% ~~~~$P_{S,G}$\dotfill &A priori eclipse prob \dotfill &$0.1960^{+0.0079}_{-0.0061}$\\
\smallskip\\\multicolumn{2}{l}{Wavelength Parameters:}&TESS\smallskip\\
~~~~$u_{1}$\dotfill &linear limb-darkening coeff \dotfill &$0.161\pm0.017$\\
~~~~$u_{2}$\dotfill &quadratic limb-darkening coeff \dotfill &$0.258\pm0.018$\\
\smallskip\\\multicolumn{2}{l}{Telescope Parameters:}&TRES\smallskip\\
~~~~$\gamma_{\rm rel}$\dotfill &Relative RV Offset$^{4}$ (m/s)\dotfill &$-1833\pm23$\\
~~~~$\sigma_J$\dotfill &RV Jitter (m/s)\dotfill &$67^{+34}_{-45}$\\
~~~~$\sigma_J^2$\dotfill &RV Jitter Variance \dotfill &$4600^{+5900}_{-4100}$\\
%\smallskip\\\multicolumn{2}{l}{Transit Parameters:}&TESS UT 2020-03-20 (TESS)&TESS UT 2020-03-27 (TESS)&TESS UT 2020-04-10 (TESS)&TESS UT 2021-12-09 (TESS)&TESS UT 2021-12-23 (TESS)&TESS UT 2022-04-03 (TESS)&TESS UT 2022-04-15 (TESS)\smallskip\\
%~~~~$\sigma^{2}$\dotfill &Added Variance \dotfill &$0.0000002169^{+0.0000000065}_{-0.0000000063}$&$0.0000001529^{+0.0000000050}_{-0.0000000048}$&$0.0000000562^{+0.0000000021}_{-0.0000000020}$&$0.0000000440^{+0.0000000022}_{-0.0000000021}$&$0.0000000570\pm0.0000000024$&$0.00000001610^{+0.00000000100}_{-0.00000000096}$&$0.00000002094^{+0.00000000098}_{-0.00000000095}$\\
%~~~~$F_0$\dotfill &Baseline flux \dotfill &$0.9999882^{+0.0000092}_{-0.0000091}$&$0.9999926^{+0.0000081}_{-0.0000082}$&$0.9999977^{+0.0000049}_{-0.0000048}$&$0.9999997^{+0.0000058}_{-0.0000059}$&$0.9999882\pm0.0000061$&$1.0000027^{+0.0000034}_{-0.0000035}$&$1.0000007\pm0.0000032$\\
\enddata
\label{exofast_table}
\tablenotetext{}{See Table 3 in \citet{Eas19} for a detailed description of all parameters}
\tablenotetext{1}{This value ignores the systematic error and is for reference only}
\tablenotetext{2}{The metallicity of the star at birth}
\tablenotetext{3}{Corresponds to static points in a star's evolutionary history. See Section 2 in \citet{Dot16}.}
\tablenotetext{4}{Reference epoch = 2460208.289100}
\tablenotetext{5}{Time of conjunction is commonly reported as the "transit time"}
\tablenotetext{6}{Time of minimum projected separation is a more correct "transit time"}
\tablenotetext{7}{Optimal time of conjunction minimizes the covariance between $T_C$ and Period}
\tablenotetext{8}{Assumes no albedo and perfect redistribution}
\end{deluxetable*}



\end{document}